\documentclass[prb,twocolumn,superscriptaddress,floatfix,showpacs]{revtex4}

\usepackage{graphicx,amsfonts,amssymb,amsmath,hyperref}

\begin{document}

\title{Algorithms for finite projected entangled pair states}

\author{Michael Lubasch}
\affiliation{Max-Planck-Institut f\"ur Quantenoptik, Hans-Kopfermann-Stra\ss{}e 1, 85748 Garching, Germany.}
\author{J.~Ignacio Cirac}
\affiliation{Max-Planck-Institut f\"ur Quantenoptik, Hans-Kopfermann-Stra\ss{}e 1, 85748 Garching, Germany.}
\author{Mari-Carmen Ba\~nuls}
\affiliation{Max-Planck-Institut f\"ur Quantenoptik, Hans-Kopfermann-Stra\ss{}e 1, 85748 Garching, Germany.}

\date{\today}

\begin{abstract}
Projected entangled pair states (PEPS) are a promising ansatz for the study of strongly correlated quantum many-body systems in two dimensions.
But due to their high computational cost, developing and improving PEPS algorithms is necessary to make the ansatz widely usable in practice.
Here we analyze several algorithmic aspects of the method.
On the one hand, we quantify the connection between the correlation length of the PEPS and the accuracy of its approximate contraction, and discuss how purifications can be used in the latter.
On the other, we present algorithmic improvements for the update of the tensor that introduce drastic gains in the numerical conditioning and the efficiency of the algorithms.
Finally, the state-of-the-art general PEPS code is benchmarked with the Heisenberg and quantum Ising models on lattices of up to $21 \times 21$ sites.
\end{abstract}


\maketitle

\section{Introduction}

As an ansatz for the wave function of a quantum many-body system, projected entangled pair states (PEPS) \cite{CiracOriginalPEPS} represent the natural generalization of matrix product states (MPS) to higher dimensions.
MPS are by now well established as a numerical tool: they constitute the variational class on which the density matrix renormalization group (DMRG) is based upon \cite{WhiteDMRG1, SchollwoeckAnn} and, nowadays, DMRG is considered numerically exact for systems comprising hundreds of quantum particles in one dimension.

PEPS have the potential to reproduce this success in higher dimensions.
This is particularly interesting for problems that cannot be tackled otherwise, e.g.\ two-dimensional fermionic or frustrated systems where quantum Monte Carlo methods are hampered by the notorious sign problem.
First PEPS results for such problems in condensed matter as the $t-J$ model or the Kagome antiferromagnet compare well to the best currently known results achieved by other means \cite{CorbozFermions2, CorboztJ, XiangKagome}.
However, the significantly higher computational cost of PEPS algorithms compared to MPS restricts the feasible simulations to PEPS tensors with much smaller dimensions, and the results are still far from the level of convergence attained in the one-dimensional case.

In the last years, significant conceptual and algorithmic progress has been made, e.g.\ Refs.~\cite{NaveTRG, XiangSU, CiraciPEPS, OrusCTM, XiangSRG, CorbozfPEPS, WangMCTNS, PizornSL, XiangHOSVD}.
Many of the numerical studies have focussed on systems in the thermodynamic limit, for which the iPEPS \cite{CiraciPEPS} ansatz can be used.
In such case, the translational invariance of the system is exploited to reduce the number of variational parameters to the few tensors in a small unit cell.
But the non translationally invariant finite PEPS ansatz is also of great importance.
On the one hand, by avoiding a predefined unit cell it allows a more unbiased approach to the thermodynamic limit, when combined with finite size scaling (although, also, a systematically increased unit cell in iPEPS can be expected to produce more and more unbiased results \cite{CorboztJ, XiangKagome}).
On the other, it is the proper ansatz for problems that are intrinsically non translationally invariant, such as the simulation of current optical lattice experiments that are being carried out in inhomogeneous traps.
In exchange, the price to pay is a more involved implementation and longer running times that scale with the system size.

The original PEPS algorithms \cite{CiracOriginalPEPS, CiracReview1} can cope with the non translationally invariant situations \cite{Murg1, Murg2, Lubasch1, SaberiKagomePEPS}, but a straightforward implementation attains only small tensor dimensions, and is not enough to explore the power of the ansatz.
In order to reach larger dimensions (i.e.\ comparable to those used in present iPEPS calculations) and to approach the optimal ground state approximations for them, it is necessary to take into account and to optimize the cost and stability of every stage of the algorithms, which is only feasible through a thorough understanding of the various possibilities.
Only then it will be possible to adopt the optimal strategies for the particular problem at hand.

In this paper we aim at a global understanding of the algorithmic aspects of finite PEPS, both at the physical and technical level.
We address the two fundamental ingredients of PEPS algorithms, namely the environment approximation, i.e.\ the approximate contraction of the tensor network (TN), and the tensor update.
In a previous article \cite{Lubasch2}, we focussed on the environment approximation.
We studied the physical significance and limitations of various contraction strategies, and introduced the cluster scheme, which unifies previous methods and gives rise to a new contraction algorithm with a trade-off between precision and computational cost.
Here, we extend the analysis of the environment approximation to provide new insight into the convergence of the cluster strategy by relating it to the correlation length of the system.
Additionally, we show how the environment approximation can be kept exactly positive with the help of purifications.
Regarding the tensor update, we investigate the effect of restricting the variational parameters to a reduced tensor \cite{CorbozfPEPS}, a technique used often in the case of iPEPS and characterized by a lower computational cost.
Via the reduced tensor we derive new numerical methods, namely suitable gauge choices, which significantly enhance the stability of the update algorithm.
These gauge choices admit also a generalization to cases where the full tensor needs to be updated.

Furthermore, we benchmark the state-of-the-art finite PEPS algorithms using the Heisenberg Hamiltonian and the quantum Ising model with transverse field.
By presenting converged finite PEPS results for lattice sizes typically considered in the context of finite size scaling, we not only assess the validity of the ansatz, but enable a systematic comparison to other methods and implementations.

The rest of the article is structured as follows.
In Sec.\ \ref{sec:prelim} we briefly present the basic notation and concepts common to PEPS algorithms.
The algorithmic details regarding the convergence of the cluster scheme, the use of positive environments and the strategies to improve the tensor update are discussed in Sec.\ \ref{sec:impr}.
Section \ref{sec:results} collects the numerical results corresponding to our best PEPS ground state approximations for the benchmark models.
Finally in Sec.\ \ref{sec:conclusions} we summarize our conclusions.

\section{Notation and preliminary concepts}
\label{sec:prelim}

For completeness we introduce here the main concepts that will be used throughout this paper.
Reviews on PEPS and more general TN methods can be found in the literature \cite{CiracReview1, OrusAnn}.

\begin{figure}
\centering
\includegraphics[width=0.35\textwidth]{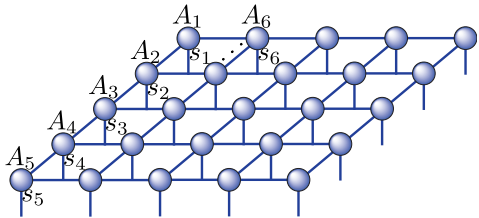}
\caption{\label{fig:PEPS}
                A PEPS on a $5 \times 5$ square lattice.
                Physical indices are depicted pointing down, and virtual ones connect tensors in the plane.
               }
\end{figure}

Given a quantum system of $N$ particles, with dimensions $d_{l}$ and local Hilbert space bases $\{|s_{l}\rangle\}_{s_{l}=1}^{d_{l}}$ (for $l=1,\ldots N$), a PEPS \cite{CiracOriginalPEPS} is a state of the form
\begin{eqnarray*}
 |\psi^{PEPS}\rangle := \sum_{s_{1}, s_{2}, \ldots, s_{N}} \mathcal{F}(A_{1}^{s_{1}} A_{2}^{s_{2}} \ldots A_{N}^{s_{N}}) |s_{1} s_{2} \ldots s_{N}\rangle,
\end{eqnarray*}
where $\mathcal{F}$ denotes the contraction of a TN formed by the tensors $A_{l}^{s_{l}}$.
Figure \ref{fig:PEPS} shows the two-dimensional square lattice geometry with open boundary conditions and size $N = L \times L$ considered throughout this work.
In this geometry, to each lattice site $l$ corresponds a tensor $A_{l}^{s_{l}}$ with one physical index $s_{l}$ for its physical degree of freedom and up to four virtual indices connecting neighboring tensors.
The dimension of the virtual indices, called \emph{bond dimension} $D$, restricts the maximum possible block entropy of the state according to an area law.

PEPS algorithms for finding ground states can be classified in two types, namely variational minimization of energy and imaginary time evolution.
With only minor changes, the second one allows also the simulation of real time evolution.
Both kinds of algorithm can be formulated in terms of the minimization of a certain cost function by varying the tensor parameters
\footnote{We remark that the initial time evolution method \cite{VidalCanForm, DaleyTE, VidalTE} works slightly different, and the basis of this work is rather given by Ref.~\cite{VerstraeteMPDO}.}.
This minimization is realized in practice by means of an alternating least squares (ALS) scheme, in which one sweeps over the tensors and updates them one after another, each time choosing the components that minimize the cost function under the constraint that all the other tensors are fixed.

Throughout this article we focus (almost exclusively) on the imaginary time evolution.
In this case, it is customary to use a Suzuki-Trotter approximation of the evolution operator where the Hamiltonian is split into parts containing only mutually commuting terms.
The cost function to be minimized is then the distance \mbox{$\mathrm{d}(|\psi\rangle) = |||\psi\rangle - \hat{O}|\phi\rangle||^{2}$}, where $|\phi\rangle$ is the initial PEPS, $\hat{O}$ is an operator representing one (or more) \emph{Trotter gates}, i.e.\ the exponential of one (or several) such Hamiltonian terms \cite{CiracReview1}, and $|\psi\rangle$ is the resulting PEPS.
During the ALS sweeping, the tensor for site $l$ is the one that minimizes
\begin{eqnarray}\label{eq:costFunction}
 \mathrm{d}(A_{l}) & =: & \vec{A}_{l}^{\dag} N_{l} \vec{A}_{l} - \vec{A}_{l}^{\dag} \vec{b}_{l} - \vec{b}_{l}^{\dag} \vec{A}_{l} + \mathrm{const}   .
\end{eqnarray}
It is given by the solution of the linear system of equations $N_{l} \vec{A}_{l} = \vec{b}_{l}$, i.e.\ $\vec{A}_{l} = N_{l}^{-1} \vec{b}_{l}$, where the norm matrix $N_{l}$ results from the norm TN $\langle \psi | \psi \rangle$ by leaving out the tensor $A_{l}^{*}$ in the bra and $A_{l}$ in the ket, and $\vec{b}_{l}$ results from the TN $\langle \psi | \hat{O} | \phi \rangle$ by leaving out $A_{l}^{*}$ in the bra.
This procedure can be iterated for the necessary number of steps to reach the desired total (real or imaginary) time
\footnote{
In the direct minimization of the energy, the cost function to be minimized is $E(|\psi\rangle) = \langle \psi | \hat{H} | \psi \rangle / \langle \psi | \psi \rangle$ over the PEPS $|\psi\rangle$.
For an update during the ALS sweeping, the tensor at position $l$ is set to $A_{l}$ minimizing $E(\vec{A}_{l}) = \vec{A}_{l}^{\dag} H_{l} \vec{A}_{l} / \vec{A}_{l}^{\dag} N_{l} \vec{A}_{l}$, which can be found as the lowest eigenstate of the generalized eigenproblem $H_{l} \vec{A}_{l} = \lambda N_{l} \vec{A}_{l}$.
Here, the matrix $H_{l}$ is defined from the TN $\langle \psi | \hat{H} | \psi \rangle$, in which one leaves out the tensor $A_{l}$ in the ket and $A_{l}^{*}$ in the bra, and $N_{l}$ is the norm matrix.
}.

Two main parts, namely the \emph{environment approximation} and the \emph{tensor update}, constitute the building blocks of this algorithm and will be often referred to in the rest of this article.
The first notion corresponds to the exact or approximate evaluation of the effective matrix ($N_{l}$) and vector ($\vec{b}_{l}$) that determine the local equation to be solved for the tensor at a given site.
The second term denotes the solution of the vector equation and the corresponding change of the PEPS with the updated tensor.

Some strategies developed in the context of iPEPS can also be applied to the finite case, and we will do that in the following.
The most widely used iPEPS method, due to its efficiency and stability, is the Simple Update (SU) \cite{XiangSU}, in which the environment is assumed to be separable and then the tensors are updated via simple SVD.
As we showed in Ref.\ \cite{Lubasch2}, the SU works equally with finite PEPS but produces results with limited accuracy.
The Full Update (FU) \cite{CiraciPEPS, CorbozfPEPS} is based on a more accurate approximation of the environment, in closer analogy to the original finite PEPS algorithm \cite{CiracOriginalPEPS, CiracReview1}, but differing from it in the fact that Trotter gates are not applied simultaneously, so that the environment for the update of one gate does only require the norm contraction around that gate.
We will in the following use the term FU in the context of finite PEPS to denote the sequential application of Trotter gates together with the full contraction of the norm TN, as in Ref.\ \cite{Lubasch2}.

\section{Algorithmic aspects}
\label{sec:impr}

In this section we analyze several distinct aspects of finite PEPS algorithms, regarding both the environment approximation and the tensor update.

In particular, for the environment approximation we show how the success of the cluster scheme introduced in Ref.\ \cite{Lubasch2} is deeply connected to the correlation length of the state.
We also discuss the feasibility and the cost of explicitly keeping a positive environment by making use of purification MPO.

For the tensor update we propose gauge choices for each possible update scheme, and show how they improve the numerical stability of the algorithms.
We additionally discuss how the reduced tensor, originally introduced in the context of iPEPS \cite{CorbozfPEPS}, can similarly be used in the finite case to speed up the computations.
The normalization of the tensors is another factor that can improve the stability of the method.

While part of this section is significantly technical, the considerations exposed here are relevant for the implementation of any (finite) PEPS algorithm.
Furthermore they have also clear physical implications, especially in the case of the environment contraction.

\subsection{Environment approximation}
\label{sec:env}

In the imaginary time algorithm, the update of one tensor at lattice site $l$ involves the contraction around that tensor of the norm TN $\langle \psi | \psi \rangle$ and of the TN $\langle \psi | e^{-\tau \hat{H}_{x}} | \phi \rangle$ for a certain subset of (mutually commuting) Hamiltonian parts $\hat{H}_{x}$.
The first contraction leads to the norm matrix $N_{l}$ and the second to the vector $\vec{b}_{l}$ from which the new tensor for that lattice site follows as $\vec{A}_{l} = N_{l}^{-1} \vec{b}_{l}$.
The original algorithm \cite{CiracOriginalPEPS, CiracReview1} includes the complete $e^{-\tau \hat{H}_{x}}$, i.e.\ all (mutually commuting) Trotter gates, in the TN $\langle \psi | e^{-\tau \hat{H}_{x}} | \phi \rangle$ and thus requires two independent environment approximations, one for $\langle \psi | \psi \rangle$ and one for $\langle \psi | e^{-\tau \hat{H}_{x}} | \phi \rangle$.
However, in the following we adopt the strategy from Ref.~\cite{Lubasch2}:
If Trotter gates are applied one by one, and only the tensors on which a given gate acts are modified, then it suffices to consider the environment approximation of the norm TN alone, and, starting from this environment, the vector $\vec{b}_{l}$ is constructed from the exact contraction of a single Trotter gate
\footnote{In an efficient implementation of this algorithm we only store and update the boundary MPO for the rows and columns of the norm TN, and the contraction for the vector $\vec{b}_{l}$ is performed on the fly.}.

\begin{figure}
\centering
\includegraphics[width=0.45\textwidth]{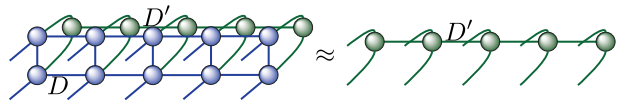}
\caption{\label{fig:BoundaryMPO}
               Original contraction \cite{CiracOriginalPEPS, CiracReview1} of the norm TN for a PEPS with bond dimension $D$.
               The product of a bulk row with a boundary MPO of bond dimension $D'$ is approximated by a new boundary MPO of the same dimension, at a cost $\mathcal{O}(dD^{6}D'^{2})+\mathcal{O}(D^{4}D'^{3})$.
              }
\end{figure}

As in the original algorithm \cite{CiracOriginalPEPS, CiracReview1}, we can approximate the environment of a PEPS row (column) with the help of \emph{boundary MPO}.
By identifying two opposite sides of the PEPS TN with boundary MPO, the action of intermediate rows (columns) on those is successively approximated by new boundary MPO, as shown in Fig.~\ref{fig:BoundaryMPO} for the norm, until the tensors of interest are reached.
The approximation accuracy of this method, used also in the FU and the cluster scheme, is ultimately determined by the \emph{boundary bond dimension} $D'$ of the boundary MPO, and its efficiency is dictated by the leading computational cost $\mathcal{O}(dD^{6}D'^{2})+\mathcal{O}(D^{4}D'^{3})$.
In typical calculations, the boundary bond dimension for a certain approximation precision scales as $D' \propto D^{2}$ independent of the system size, such that the original contraction \cite{CiracOriginalPEPS, CiracReview1} has the overall cost $\mathcal{O}(D^{10})$.

\subsubsection{Accuracy of the cluster contraction}
\label{subsec:CU}

The Cluster Update (CU) introduced in Ref.\ \cite{Lubasch2} allows a trade-off between precision and efficiency in the environment approximation.
The SU and the FU are special cases of this procedure, which naturally interpolates between them in both accuracy and computational cost.
Because clusters are not only useful for the tensor update but equally for the computation of expectation values, they realize a unifying framework for PEPS contractions.

\begin{figure}
\centering
\includegraphics[width=0.45\textwidth]{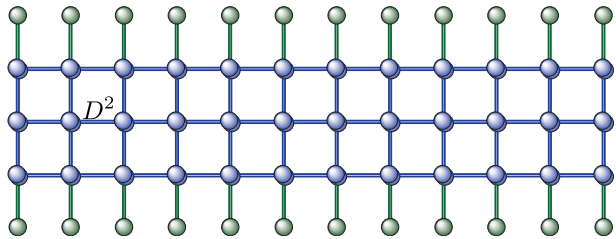}
\caption{\label{fig:ClusterSize1}
                Cluster environment of a bulk row for cluster size \mbox{$\delta=1$}.
                The contraction outside the cluster is achieved by means of a separable positive boundary MPO in the way explained in Ref.\ \cite{Lubasch2} with computational cost $\mathcal{O}(dD^{5})$.
                Then the cluster is contracted with the help of a general boundary MPO of bond dimension $D'$, which, in this case, can be found with $\mathcal{O}(dD^{5}D'^{2})$ operations.
               }
\end{figure}

A cluster is defined as a set of tensors comprising the considered ones and their neighborhood up to a distance called \emph{cluster size} $\delta$.
The idea is to approximate the environment outside the cluster very roughly and inside with more precision.
In the context of finite PEPS algorithms, in which the TN is contracted row by row, it is reasonable to define a cluster as the considered row and its neighboring rows up to the distance $\delta$.
Figure \ref{fig:ClusterSize1} shows an example cluster of size $\delta=1$ around a central row in the bulk of a PEPS.
We employ a separable positive boundary MPO for the contraction outside the cluster, which has a cost $\mathcal{O}(dD^{5})$, while the cluster itself is contracted using a general boundary MPO of bond dimension $D'>1$, which in this case requires $\mathcal{O}(dD^{5}D'^{2})$ operations \cite{Lubasch2}.
The precision and efficiency of the approximation are determined by the cluster size and $D'$.
The separable positive boundary MPO produces for cluster size $\delta=0$ an environment approximation equivalent to the SU one \cite{Lubasch2}.

We previously observed \cite{Lubasch2} that the contraction error decreases exponentially with the cluster size for PEPS ground state approximations of the Heisenberg model.
This property, which justifies the usability of clusters, is ultimately related to a finite correlation length of the system, as we appreciate here with the help of the quantum Ising model.
This model becomes critical in the thermodynamic limit at transverse field $B \approx 3.044$, and, thus, by varying $B$ we can create states with different correlation lengths.

\begin{figure}
\centering
\includegraphics[width=0.452\textwidth]{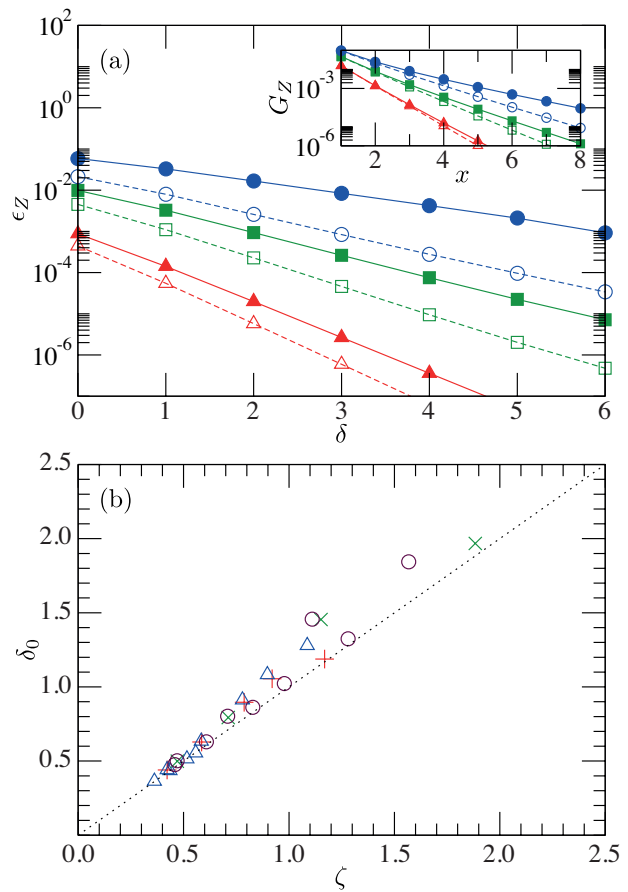}
\caption{\label{fig:ClusterCorrLength}
                Relation between cluster error and correlation function in the Ising model on a $21 \times 21$ lattice.
                (a) Cluster error (main plot) and correlation function (inset) for observable $\sigma^{Z}$, for $D=2$ (open symbols), $3$ (filled symbols), and $B=2.0$ (triangles), $2.5$ (squares), $2.8$ (circles).
                (b) Characteristic cluster size, $\delta_{0}$, versus correlation length, $\zeta$, for several values of $B \in [2,\,4]$, for observable $\sigma^{Z}$ with $D=2$ (plusses), $3$ (crosses), and for $\sigma^{X}$ with $D=2$ (triangles), $3$ (circles).
               }
\end{figure}

We have analyzed the cluster contraction error of a local observable acting on the center of the lattice, \mbox{$\epsilon_{\alpha}(\delta) := |\langle \sigma^{\alpha} \rangle_{\delta} - \langle \sigma^{\alpha} \rangle| / |\langle \sigma^{\alpha} \rangle|$}, for $\alpha=X,\,Z$, where $\langle \sigma^{\alpha} \rangle_{\delta}$ is the approximated contraction using cluster size $\delta$, and $\langle \sigma^{\alpha} \rangle$ the result of contracting the full TN.
The behavior of this quantity can be compared to the correlation function, $G_{\alpha}(x) := \langle \sigma_{l}^{\alpha} \sigma_{l+x}^{\alpha} \rangle - \langle \sigma_{l}^{\alpha} \rangle \langle \sigma_{l+x}^{\alpha} \rangle$, for two sites separated by a distance $x$ along the central column of the lattice.
All contractions were performed with large enough $D' = 100$ such that the contraction error was independent of $D'$.
\footnote{The dependence of the cluster contraction on $D'$ was already investigated previously \cite{Lubasch2}.}
We observe in Fig.~\ref{fig:ClusterCorrLength} (a) that the decrease of the contraction error is always steeper for a faster decaying correlation function.
In order to make this statement more precise, we can fit the decay of the error to an exponential function of the cluster size, $\epsilon_{\alpha}(\delta) \propto \exp(-\delta/\delta_{0})$, and obtain a characteristic cluster size $\delta_{0}$.
Correspondingly, we can extract a correlation length $\zeta$ from a similar fit of the correlation function $G_{\alpha}(x) \propto \exp(-x/\zeta)$.
After having calculated $\delta_{0}$ and $\zeta$ for several PEPS
\footnote{We determine $\delta_{0}$ via the two values of $\epsilon_{\alpha}(\delta)$ at $\delta=2$ and $4$, and $\zeta$ via the two values of $G_{\alpha}(x)$ at $x=4$ and $8$.},
we plot $\delta_{0}$ as a function of $\zeta$ in Fig.~\ref{fig:ClusterCorrLength} (b) and conclude that $\delta_{0} \approx \zeta$.
\footnote{The slight deviation of Fig.~\ref{fig:ClusterCorrLength} (b) from the exact diagonal for larger $\zeta$ is due to the fact that they correspond to $B \approx 3.0$ where both $\epsilon_{\alpha}(\delta)$ and $G_{\alpha}(x)$ decrease rather polynomially and thus $\delta_{0}$ as well as $\zeta$ depend more strongly on the two values from which they were determined.}
This demonstrates an extremely clear quantitative connection between the cluster contraction error for a given cluster size and the correlations in the state.

\subsubsection{Positive environment}
\label{subsec:posenv}

\begin{figure}
\centering
\includegraphics[width=0.308\textwidth]{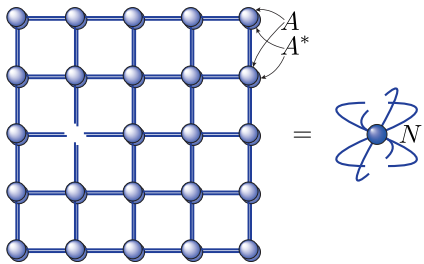}
\caption{\label{fig:PosExactNorm}
                Norm environment for a single site of a $5 \times 5$ PEPS.
                Because each PEPS tensor $A$ from Fig.~\ref{fig:PEPS} is contracted with its complex conjugate $A^{*}$ over the physical index, an exact contraction of this norm TN would give a positive hermitian norm matrix $N$.
                For large PEPS, an exact contraction of $N$ is not feasible, and the original contraction approximation \cite{CiracOriginalPEPS, CiracReview1} based on general boundary MPO, shown in Fig.~\ref{fig:BoundaryMPO}, does not keep the positivity.
               }
\end{figure}

The exact norm environment, resulting from an exact contraction of $\langle \psi | \psi \rangle$ around one (or several) site(s), is positive by construction, as can be seen in Fig.~\ref{fig:PosExactNorm}.
Although this positive characteristic is considered a desirable property for the environment approximation, in general it is not respected by the approximated contractions.
Nevertheless, it is possible to use schemes that maintain it.
In particular, the Single-Layer (SL) algorithm was introduced in Ref.\ \cite{PizornSL} as a way to improve the environment approximation of the SU while preserving its efficiency and numerical stability.
The SL method performs the norm contraction by means of transformations in the ket alone \cite{PizornSL}.
Then the boundary is described by a \emph{purification} MPO \cite{VerstraeteMPDO}, defined via a MPS of virtual bond dimension $D''$ and physical dimension $D \times d'$ in such a way that the MPO results from tracing over the \emph{purification bonds} of dimension $d'$.
Approximating the environment in the SL way and then updating the tensors as explained in Ref.\ \cite{PizornSL} ensures a stable algorithm, but as seen in Ref.\ \cite{Lubasch2} the error in the environment approximation can be several orders of magnitude above that of the original contraction \cite{CiracOriginalPEPS, CiracReview1}, and it can depend strongly on the system size.

Several factors can cause these accuracy limitations.
Even if  there exists a good positive MPO approximation for the boundary with moderate bond dimension $D'$ (as observed for gapped systems \cite{CiracBoundaryTheories}), it does not necessarily follow that $D''$ is small \cite{Gemma, Kliesch} and hence it is not clear a priori that fixing the maximum $D''$ produces an accurate approximation for the environment.
Moreover, as argued in Ref.\ \cite{Lubasch2}, by operating on a single layer, the scheme does not find the most general purification with given bond $D''$.
Here we want to address the question wether the accuracy limitations of the SL algorithm are due to the description of the boundary as purification or wether they are due to the specific operations proposed in Ref.~\cite{PizornSL} to determine that boundary purification.

\begin{figure}
\centering
\includegraphics[width=0.458\textwidth]{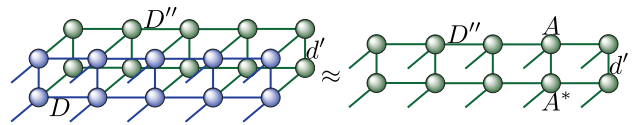}
\caption{\label{fig:Purification}
               Positive contraction of the norm TN for a PEPS with bond dimension $D$. 
               The product of a bulk row with a boundary purification of virtual bond dimension $D''$ and purification bond dimension $d'$ is approximated by a new boundary purification of the same dimensions.
               Each update of a tensor $A$ constitutes a nonlinear problem, and solving the linearized equations costs $\mathcal{O}(dD^{6}D''^{4})+\mathcal{O}(D^{4}D''^{6})+\mathcal{O}(d'^{3}D^{3}D''^{6})$ where $d' \leq DD''^{2}$.
              }
\end{figure}

One way to allow for a more general purification is to formulate an algorithm in the double-layer picture in the following way.
Given the MPS bond dimension, $D''$, and the purification bond, $d'$, we write a purification MPO \cite{VerstraeteMPDO} by using in the lower layer the complex conjugated tensors from the upper layer.
The problem of approximating the boundary after the contraction of one further row of the PEPS norm TN is then formulated for this structure instead of the general MPO, as sketched in Fig.~\ref{fig:Purification}.
The local equations result from replacing the single tensor of the general MPO by the structure consisting of $A_{l}$ and $A_{l}^{*}$.
Following the standard ALS procedure, we sweep over the sites $l$, and for each site solve the corresponding optimization problem for $A_{l}$.
However, in this case the cost function to be minimized is no longer quadratic, but quartic in the variables of a tensor at site $l$, and its minimum corresponds to the solution of nonlinear equations, in contrast to the linear equations encountered in the original contraction \cite{CiracOriginalPEPS, CiracReview1} of Fig.~\ref{fig:BoundaryMPO}.
The nonlinear equations for $A_{l}$ have to be solved iteratively.
We describe and benchmark several options in Appendix \ref{app:puri}.

\begin{figure}
\centering
\includegraphics[width=0.445\textwidth]{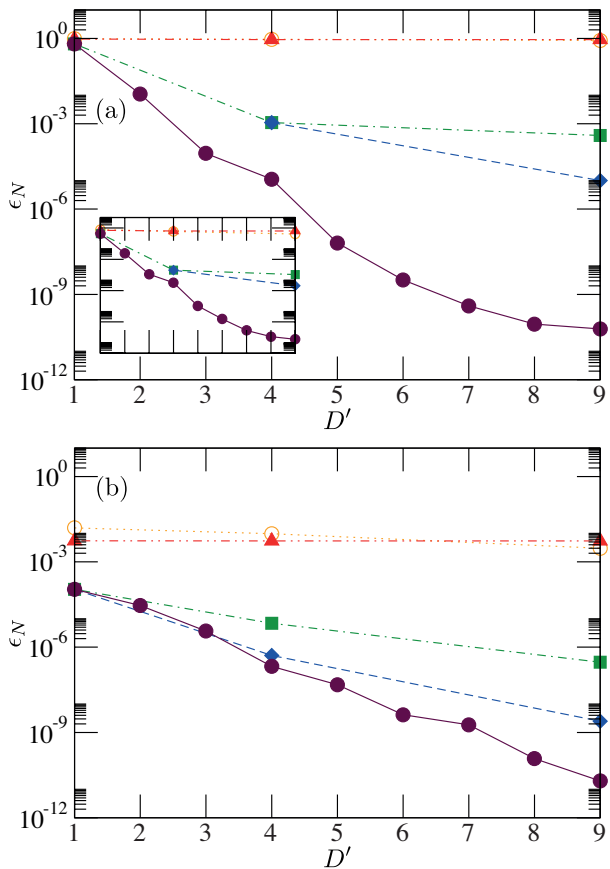}
\caption{\label{fig:NormErrorPuri}
                Relative error of the norm contraction using boundary purification MPO, for SU Ising ground state approximations.
                (a) $B=3.0$ and $D=2$ on lattices of size $21 \times 21$ (main plot) and $11 \times 11$ (inset).
                (b) $B=1.0$ and $D=4$ on a $11 \times 11$ lattice.
                For reference, we show the error of the SL method with maximum $d'=DD''^{2}$ (open circles) and of the original algorithm \cite{CiracOriginalPEPS, CiracReview1} (filled circles).
                Our purification contraction was performed with $d'=1$ (triangles), $2$ (squares), and $3$ (diamonds), and $D''=\sqrt{D'}$.
              }
\end{figure}

We compare this scheme to the original \cite{CiracOriginalPEPS, CiracReview1} and the SL algorithm based on the norm contraction of the same PEPS used in the previous analysis of Ref.\ \cite{Lubasch2}.
The results are shown in Fig.~\ref{fig:NormErrorPuri}.
For a fixed purification bond $d'$, we observe that the relative error of the norm decreases fast as a function of $D''$.
The comparison of the $11 \times 11$ to the $21 \times 21$ lattice (Fig.~\ref{fig:NormErrorPuri} (a)) shows that, similar to the original algorithm \cite{CiracOriginalPEPS, CiracReview1}, the error does not have a strong dependence on the system size.
On the other hand, with growing $d'$, the curves tend to converge to the error of the original contraction \cite{CiracOriginalPEPS, CiracReview1}.
This effect can be observed already with small purification bonds for $B=1.0$ (Fig.~\ref{fig:NormErrorPuri} (b)).
These results suggest that the error in the SL method is mainly due to the restricted class of purifications it can attain, and not to the description of the boundary as a purification with small bond $D''$.
\footnote{Although our conclusions are based on PEPS from the SU, because we had thoroughly analyzed exactly the same PEPS with the SL algorithm in Ref.~\cite{Lubasch2}, we clearly expect similar improvements for PEPS from the FU (in particular the ground state of the quantum Ising model at $B = 1.0$, away from criticality, should be well approximated by the SU and thus Fig.~\ref{fig:NormErrorPuri} (b) should only change slightly for the corresponding FU PEPS).}

From the discussion above we conclude that it is possible to efficiently find a (close to) optimal general purification by means of the solution of nonlinear equations.
In the context of PEPS contractions, this technique improves the SL scheme significantly, but given its higher computational cost compared to the original contraction \cite{CiracOriginalPEPS, CiracReview1}, resulting from the iterative routines (see Appendix \ref{app:puri}), it is not a practical option.
Hence, in the following, all our cluster and full contractions will be based upon the original contraction algorithm \cite{CiracOriginalPEPS, CiracReview1} and thus make use of general boundary MPO as shown in Fig.~\ref{fig:BoundaryMPO}.
Nevertheless, the procedures analyzed here may be useful for other problems where the question of numerically optimizing a purification MPO appears, such as for the description of one-dimensional thermal states or open systems.

\subsection{Tensor update}
\label{sec:update}

Once the environment is computed, the actual update of the tensors takes place by solving the appropriate local equations.
It is also possible to use simplifications of this step which render a more efficient and stable algorithm.

\begin{figure}
\centering
\includegraphics[width=0.15\textwidth]{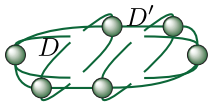}
\caption{\label{fig:EnvTensorPair}
                The $6$ environment tensors of a nearest-neighbor tensor pair.
                They form a periodic boundary MPO with virtual bond dimension $D'$ and physical dimension $D$.
                It is constructed with $\mathcal{O}(dD^{6}D'^{2})+\mathcal{O}(D^{4}D'^{3})$ operations, resulting from the optimal search for a boundary MPO and the contraction of the environment up to the tensor pair.
               }
\end{figure}

For the update of a pair of neighboring tensors, the environment can be approximated in general by a MPO with periodic boundary conditions, as illustrated in Fig.~\ref{fig:EnvTensorPair}.
A first simplification of the tensor update procedure comes from sequentially processing the Trotter gates, as described above, and changing only the tensors on which each gate acts.
Then all the update operations on the pair are performed with a fixed environment.

The computational cost of the tensor update can be greatly decreased by restricting the update to the \emph{reduced tensor} \cite{CorbozfPEPS}.
This reduced tensor update minimizes the cost function \eqref{eq:costFunction} for the full tensors exactly only in the case of a separable environment \cite{Lubasch2}, as e.g.\ in the SU, but it is worth studying its performance in a more general situation.
In any case, it allows to work with larger bond dimensions, which might compensate for the smaller number of variational parameters.

Another major difference between MPS and PEPS concerns the conditioning of the effective norm matrix $N_{l}$.
For MPS with open boundary conditions, a \emph{gauge transformation}
\footnote{
In a PEPS, for any pair of neighboring tensors that are connected via a virtual index, an arbitrary matrix $M$ can be contracted with one tensor and the matrix $M^{-1}$ with the other in such a way that the state does not change.
This establishes the analogue of a (local) gauge freedom.
Throughout this article, the matrix $M$ is called \emph{gauge matrix} or \emph{gauge transformation} in accordance with Ref.~\cite{CiracReview1}, and the term \emph{gauge fixing} or \emph{gauge choice} refers to the process of choosing a specific matrix $M$.
}
can be chosen such that $N_{l}=1$, which guarantees the stability of the tensor update.
Although this is impossible for PEPS, we will show how a proper gauge choice and tensor normalization drastically improve the stability of the algorithm.

\subsubsection{Reduced tensor}

\begin{figure}
\centering
\includegraphics[width=0.421\textwidth]{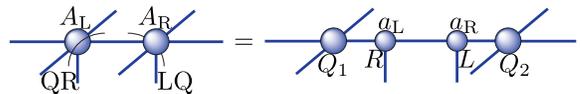}
\caption{\label{fig:RedTensor}
                A QR decomposition of the left full tensor $A_{\mathrm{L}}$ generates the left reduced tensor $a_{\mathrm{L}}$ as the $R$.
                Similarly, a LQ decomposition of the right full tensor $A_{\mathrm{R}}$ gives the right reduced tensor $a_{\mathrm{R}}$ as the $L$.
                The initial $dD^{4}$ variational parameters of the full tensor are decreased to the $d^{2}D^{2}$ variational parameters of the reduced tensor.
               }
\end{figure}

Before performing the update under a nearest-neighbor Trotter gate, the tensor for a lattice site can be decomposed into the contraction of two tensors, in such a way that one of them carries the physical index and the virtual bond corresponding to the link on which the two-site gate acts, i.e.\ all the indices directly affected by the gate.
This tensor is called the \emph{reduced tensor} \cite{CorbozfPEPS}, and can be obtained from the full tensor by means of a QR decomposition, as sketched in Fig.~\ref{fig:RedTensor}.

\begin{figure}
\centering
\includegraphics[width=0.405\textwidth]{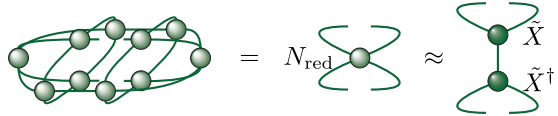}
\caption{\label{fig:EnvRedTensorPair}
                Environment tensor $N_{\mathrm{red}}$ of a reduced nearest-neighbor tensor pair and its closest positive semidefinite approximant $\tilde{X} \tilde{X}^{\dag}$ constructed as explained in the text.
                The contractions are characterized by the leading computational cost $\mathcal{O}(d^{4}D^{4}D'^{2})+\mathcal{O}(d^{2}D^{6}D'^{2})+\mathcal{O}(d^{2}D^{4}D'^{3})$ and the computation of the positive approximant requires additionally
                $\mathcal{O}(d^{6}D^{6})$ operations.
               }
\end{figure}

In the reduced tensor update, only the components of such reduced tensor are modified during the update procedure, while the remaining part of the full tensor is left unchanged.
These remaining parts of both tensors in the pair are contracted with the periodic MPO of Fig.~\ref{fig:EnvTensorPair} to get the environment for the reduced tensor pair, $N_{\mathrm{red}}$, shown in Fig.~\ref{fig:EnvRedTensorPair}.
Due to the approximate contractions, this reduced environment is in general not positive, neither is it hermitian, but its positive approximant can be constructed in two steps \cite{Keller}.
First, we compute the optimal hermitian approximant $\tilde{N}_{\mathrm{red}}:=(N_{\mathrm{red}}+N_{\mathrm{red}}^{\dag})/2$.
Second, from its eigendecomposition $\tilde{N}_{\mathrm{red}} = U \Sigma U^{\dag}$ we obtain the positive approximant as $U \Sigma_{+} U^{\dag}$ where $\Sigma_{+}$ results from $\Sigma$ by setting all negative eigenvalues to zero.
Finally, the environment is written as $\tilde{X} \tilde{X}^{\dag}$ in terms of its square root $\tilde{X} := U \sqrt{\Sigma_{+}}$.

The computational cost of the contractions for the periodic boundary MPO (Fig.~\ref{fig:EnvTensorPair}), needed in both the reduced and the full tensor update, reads $\mathcal{O}(dD^{6}D'^{2})+\mathcal{O}(D^{4}D'^{3})$.
The construction of $N_{\mathrm{red}}$ (Fig.~\ref{fig:RedTensor} and \ref{fig:EnvRedTensorPair}) is only slightly more expensive with $\mathcal{O}(d^{4}D^{4}D'^{2})+\mathcal{O}(d^{2}D^{6}D'^{2})+\mathcal{O}(d^{2}D^{4}D'^{3})$ operations.
Its eigendecomposition requires $\mathcal{O}(d^{6}D^{6})$.
In the complete update of the reduced tensors via the sweeping of the ALS scheme, all further operations have lower computational cost.
Notice that in the case of the full tensors the contraction of the norm environment for a single tensor needs $\mathcal{O}(D^{8}D'^{2})+\mathcal{O}(D^{4}D'^{3})$ operations while the eigendecomposition of the norm matrix has the cost $\mathcal{O}(d^{3}D^{12})$.

In order to study wether the reduced tensor limits the accuracy of the method, we considered imaginary time evolution of the Heisenberg model on $4 \times 4$ and $10 \times 10$ lattices, and compared the final energies from the reduced tensor update to the ones from the full tensor update.
We found that, while for the small bond dimensions $D=2$ and $3$ the full tensor update produced better energies, for $D=4$ the energies of both approaches were already very similar.
This can be appreciated by comparison of the results in Appendix \ref{app:pepsE} (obtained with the reduced tensor) to the full tensor results published in Refs.\ \cite{CiracOriginalPEPS, Murg2}.

Because the reduced tensor update is less costly, we could reach larger bond dimensions than with the full tensor update, and, in the end, obtained the lowest energies with the reduced tensors.
Therefore the reduced tensor update was used for the results presented in this paper.

\subsubsection{Gauge fixing}
\label{subsec:gauge}

In the case of MPS, it is possible to keep up a canonical form of the tensors during their updates with the help of a local gauge fixing, and that ensures the stability of the algorithm and optimizes its performance \cite{CiracReview1}.
In the case of PEPS, there exists neither such a canonical form nor any means to locally gauge away the norm matrix.
Nevertheless, using the gauge freedom, it is possible to improve the conditioning of the norm matrix and positively affect the precision and stability of the method, as we describe in the following.
\footnote{A recent alternative approach for an improved imaginary time evolution is presented in Ref.~\cite{OrusQuasiOrth}, in which the authors propose a ``quasicanonical form'' that arises as fix point of the SU performed with nearest-neighbor identity gates, and in which they demonstrate how that form allows an efficient and stable imaginary time evolution with projected entangled pair operators (PEPO) in the context of iPEPS.}

We propose a gauge fixing that is inspired by the one-dimensional case with open boundary conditions.
In that case, the norm tensor can be reduced to the identity by (partially) imposing the canonical form of the MPS, achieved by QR (or LQ) decomposition of each tensor after its update \cite{CiracReview1}.
Alternatively, for an arbitrary MPS it is always possible to reduce the norm matrix to the identity by taking the square roots of the unconnected left and right environment halves and absorbing part of their QR (LQ) decompositions in the tensor to be updated.

In the case of PEPS, it is not possible to ensure an identity norm matrix by means of QR or LQ decompositions after the tensor update.
Hence, we adapt the second possibility and obtain the gauge transformations from the environment before the tensor update, namely from the norm tensor itself, such that the norm matrix is better-conditioned.
Because this gauge fixing can be combined with any of the environment approximations described previously, we propose a precise scheme for each case.

When the environment of the tensor pair is separable, i.e.\ $D'=1$ in Fig.~\ref{fig:EnvTensorPair}, it decomposes into six positive semidefinite matrices, which can be determined by the algorithm in Ref.\ \cite{Lubasch2}.
We compute the square roots of these matrices and absorb them in the tensor pair.
After contraction of the tensor pair with the Trotter gate, a SVD is performed to find the new tensors, and finally these are multiplied by the inverses of the previous square roots.
This procedure coincides with the SU \cite{XiangSU} in which the $\lambda$ matrices surrounding the tensor pair are substituted here by the square roots of the environment matrices corresponding to each link.
Since the positive separable environment of the tensor pair is obtained with $\mathcal{O}(dD^{5})$ operations, the leading cost of the complete update is $\mathcal{O}(d^{6}D^{3})+\mathcal{O}(d^{2}D^{5})$, under the assumption $d \leq D^{2}$.

\begin{figure}
\centering
\includegraphics[width=0.472\textwidth]{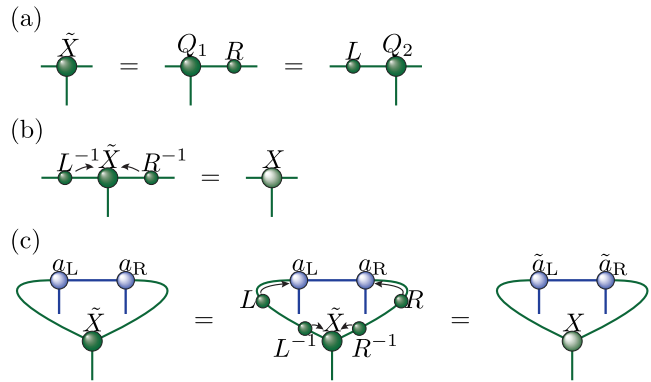}
\caption{\label{fig:GaugeEnvRedTensorPair}
                Gauge fixing on the environment tensor of the reduced tensor pair, when the environment is non-separable.
                (a) We perform a QR and LQ decomposition on $\tilde{X}$ from Fig.~\ref{fig:EnvRedTensorPair} independently of each other (notice that we have shortened here the horizontal open indices of $\tilde{X}$ compared to Fig.~\ref{fig:EnvRedTensorPair}).
                (b) Contraction of $\tilde{X}$ with $L^{-1}$ and $R^{-1}$ gives the final square root of the environment tensor, $X$.
                (c) In order to leave the state unchanged, the left and right reduced tensors $a_{L}$ and $a_{R}$ from Fig.~\ref{fig:RedTensor} have to be contracted with the gauge transformations $L$ and $R$ as shown here, which gives the starting tensors
                      $\tilde{a}_{L}$ and $\tilde{a}_{R}$ for the update explained in Fig.~\ref{fig:UpdateRedTensorPair}.
               }
\end{figure}

When the environment of the tensor pair is non-separable, and we restrict the update to the reduced tensor, we propose the gauge fixing from Fig.~\ref{fig:GaugeEnvRedTensorPair}.
By taking $R$ and $L$ from independent QR and LQ decompositions of the same $\tilde{X}$ from Fig.~\ref{fig:EnvRedTensorPair}, we treat both virtual bonds of the environment equally, such that both reduced tensors will experience similar condition numbers in the linear equations of the following sweeping
\footnote{
We also explicitly checked the resulting condition numbers when first a QR and then a LQ, or the other way round, or only a single decomposition is applied.
In these cases, one tensor always encounters better condition numbers than the other in the linear equations of the following sweeping.
}.
After we have obtained the desired better-conditioned square root of the environment tensor, $X$, in order to leave the state unchanged, the left (right) reduced tensor has to be contracted with $L$ ($R$) over its left (right) virtual index.

\begin{figure}
\centering
\includegraphics[width=0.472\textwidth]{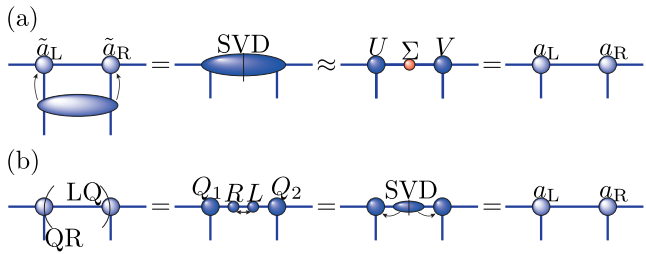}
\caption{\label{fig:UpdateRedTensorPair}
                Initial and final step of the reduced tensor update.
                (a) Initialization: Before we update the reduced tensors by sweeping, we apply a SVD to their joint contraction with the Trotter gate, and keep only the $D$ largest singular values in $\Sigma$.
                      Splitting $\Sigma$ gives the initial tensors $a_{\mathrm{L}}:=U\sqrt{\Sigma}$ and $a_{\mathrm{R}}:=\sqrt{\Sigma}V$ for the ALS procedure.
                (b) Final form: After convergence of the ALS sweeping, we put the two tensors on an equal footing.
               }
\end{figure}

After our gauge fixing has been applied, the actual update takes place in three steps.
First, the tensors are initialized using a SVD as shown in Fig.~\ref{fig:UpdateRedTensorPair} (a).
This step coincides with the SU.
If the environment is separable, the cost function is already minimal.
In any other case, we can anticipate good starting tensors that are closer to the minimum of the cost function.
Second, we optimize the tensors by means of the standard ALS sweeping, in which each tensor update is followed by the standard gauge fixing \cite{CiracReview1}, i.e.\ the left (right) tensor is QR (LQ) decomposed along its right (left) virtual bond.
Third, a gauge choice is made on the internal link of the converged pair as shown in Fig.~\ref{fig:UpdateRedTensorPair} (b).

We have observed that our gauge choices improve the condition number of the norm matrix by several orders of magnitude in all studied cases.
This statement is quantified by the results in Tab.~\ref{tab:CondnumsRedTensor}, which compares typical condition numbers found in the simulation of the Ising and Heisenberg models with and without our gauge fixing.
Strictly speaking, the condition number of the norm matrix $N_{l}$ provides only an upper bound for the final error of the solution $\vec{A}_{l}$ to the linear system of equations \cite{Golub}:
Therefore, a large condition number does not imply low accuracy, but a small condition number implies high accuracy of the solution.
In practical computations with finite PEPS, when our gauge transformations are not used instabilities can occur (e.g.\ as reported in Ref.~\cite{PizornSL}) that we have never encountered after our gauge fixing.

\begin{table}
\centering
 \begin{tabular}{ c c c c }
  \hline \hline
  (a) & Model & Positive approximant & Gauge fixing \\ \hline
        & $B=1.0$ Ising    & $(2 \pm 3) \cdot 10^{7}$ & $1.1 \pm 0.1$ \\
        & $B=3.0$ Ising    & $(2 \pm 3) \cdot 10^{3}$ & $1.6 \pm 0.1$ \\
        & Heisenberg        & $(8 \pm 5) \cdot 10$       & $1.08 \pm 0.02$ \\
  \hline \hline
 \end{tabular}
 \begin{tabular}{ c c c c }
  \hline \hline
  (b) & Model & Positive approximant & Gauge fixing \\ \hline
        & $B=1.0$ Ising    & $(9 \pm 205) \cdot 10^{13}$ & $(1 \pm 3) \cdot 10^{4}$ \\
        & $B=3.0$ Ising    & $(4 \pm 158) \cdot 10^{13}$ & $(5 \pm 6) \cdot 10^{2}$ \\
        & Heisenberg        & $(3 \pm 2) \cdot 10^{4}$        & $5 \pm 3$ \\
  \hline \hline
 \end{tabular}
\caption{\label{tab:CondnumsRedTensor}
                We show the mean condition number of the norm matrix with its standard deviation in the reduced tensor update without our gauge fixing, using only the positive approximant, and with our gauge fixing during the FU imaginary time evolution
                of $D=2$ (a) and $D=4$ (b) PEPS of size $N = 11 \times 11$ for the Ising model and of size $N = 10 \times 10$ for the Heisenberg model.
                The values were obtained averaging over $10$ time steps and all tensors in the lattice.
               }
\end{table}

\begin{figure}
\centering
\includegraphics[width=0.461\textwidth]{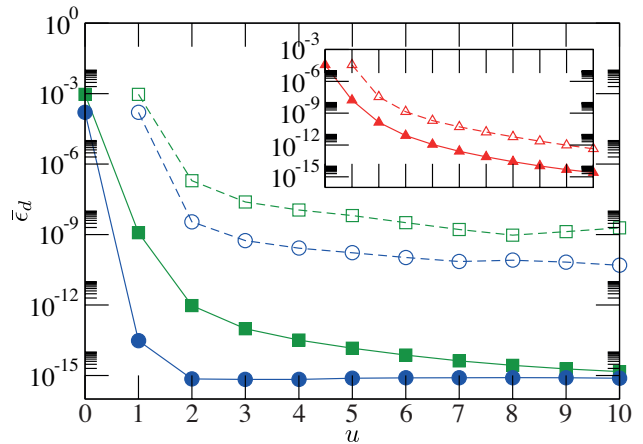}
\caption{\label{fig:GaugeErrorRedTensor}
                Mean value of the relative change \mbox{$\epsilon_{d}(u) := |d(u)-d(u-1)|/|d_{\mathrm{init}}|$} of the cost function $d$, Eq.~\eqref{eq:costFunction}, after consecutive update sweeps $u$ over a tensor pair computed with respect to the initial value
                of the cost function $d_{\mathrm{init}}$, for the $D=4$ reduced tensor update setting of Tab.~\ref{tab:CondnumsRedTensor}.
                We compare the FU evolution without our gauge fixing using only the positive approximant (open symbols) to the same propagation with our gauge fixing (filled symbols), for a $11 \times 11$ Ising model at $B=1.0$ (circles) and $3.0$ (squares),
                and for a $10 \times 10$ Heisenberg model (triangles).
               }
\end{figure}

We can also investigate the effect of our gauge fixing on the convergence of the ALS sweeping, which can be gathered from Fig.~\ref{fig:GaugeErrorRedTensor} for the update of the reduced tensor.
Most remarkably, in the presence of the gauge transformations, already the initial SVD drastically reduces the cost function Eq.~\eqref{eq:costFunction}, by a value that in all considered cases is larger than the one attained after one sweep without the gauge transformations.
Furthermore, the final total reduction of the cost function is also larger with our gauge fixing than without.
Because the relative change of the cost function in Fig.~\ref{fig:GaugeErrorRedTensor} always decreases faster in the presence of our gauge transformations, we conclude that the latter accelerate the convergence of the ALS scheme.
Our results indicate that a simplified tensor update consisting of the combination gauge fixing and SVD only, without the ALS sweeps, might be successful.
Indeed, for the Ising model, the cost function after our gauge fixing and SVD is already smaller than after $10$ sweeps without our gauge fixing.
However, the sweeping can further decrease the cost function, and this is revealed most evidently for the Heisenberg model.

So far, we assumed that the update is performed on two directly neighboring tensors, after applying on them one of the Trotter gates of a nearest-neighbor Hamiltonian.
The discussion can be extended to the update of more distant tensors, as would appear in the case of Hamiltonians with long-range interactions.
However, for $n$ non-adjacent reduced tensors the dimension of the norm matrix is typically $(d^{2}D^{2})^{n} \times (d^{2}D^{2})^{n}$, and its diagonalization, even for two tensors, is not desirable.
A further simplification is to choose the gauge transformations from the local environment of a single tensor, since the norm matrix in this case has size $d^{2}D^{2} \times d^{2}D^{2}$.
By means of a numerical simulation we confirmed that such a local gauge choice can produce condition numbers comparable to the ones obtained from the gauging of the environment of the pair
\footnote{We propagated, using the FU, several $11 \times 11$ $D=2$ PEPS, obtained from the SU for the Ising model with different fields, and we monitored eigenvalues and singular values of norm and gauge matrices throughout the evolution.}.
Moreover, we found that the gauge matrices $L$ and $R$ computed from the local norm tensors of each separate tensor in the pair can also be applied to the pair environment, and then we can follow the update procedure of Fig.~\ref{fig:UpdateRedTensorPair}.
Thus, we expect that local gauge choices similarly improve the tensor update in the case of general long-range interactions, where the tensor initialization and final form in Fig.~\ref{fig:UpdateRedTensorPair} will be given by their analogues from TEBD \cite{VidalCanForm}.

While the discussion here is focussed on the reduced tensors, in Appendix \ref{app:gaug} we derive an efficient gauge fixing for the full tensors.
This gauge fixing equally improves the condition number of the norm matrix and the convergence of the ALS sweeping in the tensor update of the full tensors.
Because all our gauge transformations are derived from and applied to the norm TN alone, they do not explicitly depend on the operator whose action on the PEPS is approximated.
We would therefore expect that our gauge choices similarly improve the original time evolution algorithm \cite{CiracOriginalPEPS, CiracReview1} in which the action of projected entangled pair operators (PEPO) on PEPS is approximated
\footnote{We can also expect that the direct variational minimization of the energy benefits from our gauge transformations because they improve the condition number of the norm matrix $N_{l}$ that enters the generalized eigenproblem $H_{l} \vec{A}_{l} = \lambda N_{l} \vec{A}_{l}$ which has to be solved for the tensor update at lattice site $l$.}.

\subsubsection{Stability issues}

The previously described gauge choices guarantee a better conditioned norm matrix.
But for the stability, precision and efficiency of the algorithms, especially when the environment approximation is very rough (e.g.\ by using small clusters or boundary bond dimensions), also the following factors need to be taken into account.

\begin{itemize}

\item
For PEPS, the matrices $N_{l}$ are not exactly hermitian and positive semidefinite.
The advisable strategy is to replace them by their closest hermitian approximants $(N_{l}+N_{l}^{\dag})/2$, and additionally set to zero any negative eigenvalues in order to get the closest positive semidefinite approximant of $N_{l}$, as described above for the environment of the reduced tensor pair,
\footnote{
In the direct variational minimization of the energy, also the matrices $H_{l}$ are not exactly hermitian and hence should be replaced by their closest hermitian approximants $(H_{l}+H_{l}^{\dag})/2$.
Furthermore, the eigendecomposition of $(N_{l}+N_{l}^{\dag})/2$ enables us to replace the generalized eigenproblem by a standard one, $\sqrt{N_{l}}^{-1} H_{l} \sqrt{N_{l}}^{-1} \vec{B}_{l} = \lambda \vec{B}_{l}$.
Its lowest eigenvector, $\vec{B}_{l}$, yields the desired new variational parameters via $\vec{A}_{l} = \sqrt{N_{l}}^{-1} \vec{B}_{l}$.
}.

\item
In general, some eigenvalues of $N_{l}$ are zero and its positive subspace is ill-conditioned.
That is why $N_{l}^{-1}$ must be a pseudoinverse.
A cutoff is set such that only the subspace of $N_{l}$ with eigenvalues larger than a certain value is considered in the construction of the pseudoinverse.

\item
Finally, the correct tensor normalization has a decisive impact.
Imaginary time evolution steadily modifies the norm of the state.
Thus we impose the normalization of the PEPS, $\langle \psi | \psi \rangle = 1$, after each set of Trotter gates, and, in order to avoid the existence of very small or very large tensors, we additionally scale all PEPS tensors to have the same largest element absolute.

\end{itemize}

\newpage

\section{Performance of finite PEPS}
\label{sec:results}

With the aim of analyzing its performance in terms of system size and bond dimension, we have applied the generic finite PEPS code to the ground state search for the Heisenberg and quantum Ising model, and compared the results to those obtained by other numerical methods, when available.
Our best PEPS results were obtained with the FU, i.e.\ updating the reduced tensors, applying the Trotter gates sequentially, and approximating the full contraction of the environment by means of general boundary MPO.
This combination of techniques allowed us to push the simulations to lattices of size up to $21 \times 21$.
On the same systems, we ran also the SU for finite PEPS.

\subsection{Convergence procedure}

In each case, the PEPS ground state approximation was found by means of imaginary time evolution.
The initial state was always a $D=2$ PEPS which was constructed by embedding in it a separable PEPS and replacing the zero entries by small random numbers.
Beginning with the time step $\tau=0.01$, the propagation was performed long enough for the energy to converge, and then the procedure was repeated for smaller time step(s).
After convergence was attained for the minimum time step, the scheme was iterated for a larger bond dimension, starting from a previously converged PEPS as initial state.

We observed that the converged SU PEPS of a certain bond dimension was always a good initial state for further propagation with the FU for this bond dimension.
On the one hand, in general, the SU PEPS can already be a good ground state approximation and then only few further steps with the FU are required.
On the other hand, we have found, that such state required smaller values of $D'$ when the evolution was continued with the FU.

Energies and correlators reported here for a certain value of $D$ correspond to the final PEPS for the smallest time step.
The error in the corresponding observable was estimated via the difference to the expectation value calculated with the converged PEPS for the previous time step.
All contractions were performed with boundary bond dimension $D'=100$, big enough to neglect contraction errors, as we explicitly checked by comparison to results from $D'=200$.

\subsection{Heisenberg model}

We considered imaginary time evolution with an antiferromagnetic Heisenberg Hamiltonian \mbox{$\hat{H} = \sum_{\langle l, m \rangle} \vec{S}_{l} \cdot \vec{S}_{m}$}.
This model on a two-dimensional square lattice is a paradigmatic benchmark Hamiltonian because quantum Monte Carlo methods provide quasi exact results for very large system sizes \cite{SandvikSSE}, and thus we can directly compare our results to quantum Monte Carlo
\footnote{We obtain our quantum Monte Carlo reference values from the ALPS library \cite{ALPS1, ALPS2, ALPS3}.}.
In the context of PEPS, the ground state order parameter of this model, i.e.\ the squared staggered magnetization $M_{\mathrm{stag}}^{2} := \frac{1}{N^{2}} \sum_{l, m = 1}^{N} (-1)^{l+m} \langle \vec{S}_{l} \cdot \vec{S}_{m} \rangle$, is particularly challenging \cite{BaueriPEPS} (also on a honeycomb lattice \cite{XiangSRG}) and a precise determination has so far only been possible with very large bond dimension $D=16$ in Ref.~\cite{WangMCTNS}.
Here we want to find out what our improved algorithmic procedures can do.

To our Heisenberg Hamiltonian we added a small staggered magnetic field $B_{Z} \sum_{l} (-1)^{l} S_{l}^{Z}$ which we slowly switched off during the evolution, starting from $B_{Z} = 10^{-3}$.
In the presence of this staggered field the SU(2) symmetry of the Heisenberg model is explicitly broken and smaller values of $D'$ suffice.
This procedure improved the convergence of all our algorithms significantly.
In the case of the SU, it helped to avoid local minima and reach lower final energies, in particular on the largest $20 \times 20$ lattice.
And, in the case of the FU, already when the staggered field was still switched on, low values of the energy were attained while smaller values of $D'$ were required.
All propagations were performed for time steps $\tau = 10^{-2}$ and $10^{-3}$.

\begin{figure}
\centering
\includegraphics[width=0.439\textwidth]{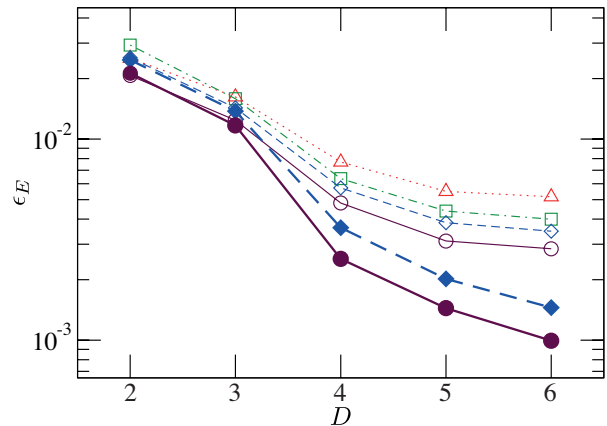}
\caption{\label{fig:HeisEnergies}
                Relative energy error $\epsilon_{E}:=|E(D)-E_{0}|/|E_{0}|$, where $E_{0}$ denotes the exact ground state energy from the ALPS library \cite{ALPS1, ALPS2, ALPS3}, of the SU (open symbols) and the FU (filled symbols) for different lattice sizes.
                In the case of the SU, we consider $N = 10 \times 10$ (triangles), $14 \times 14$ (squares), $16 \times 16$ (diamonds), and $20 \times 20$ (circles).
                In the case of the FU, we consider $N = 10 \times 10$ (diamonds) and $14 \times 14$ (circles).
               }
\end{figure}

Figure \ref{fig:HeisEnergies} shows the convergence of the energy with increasing bond dimension.
We observe that, while the FU energy error decreases rapidly with $D$, the SU energies saturate, and for bond dimensions up to $D=6$, the lowest SU energies lie between the values for $D=3$ and $4$ obtained with the FU.
This is consistent with our earlier observations in Ref.\ \cite{Lubasch2} based on smaller lattices.
Both the SU and the FU produce better energies when the lattice size increases.

We can now compare our energy accuracies to the existing literature.
The original finite PEPS algorithm \cite{CiracOriginalPEPS, CiracReview1} obtained a lowest energy per site $-0.62515$ on a $10 \times 10$ lattice, using time step $\tau = 0.001$ and bond dimension $D = 4$ (this PEPS result is given in Refs.~\cite{BanulsSGS, MezzacapoEPS}).
For this system size and the same values of $\tau$ and $D$ we now achieve the slightly lower energy per site $-0.62637(2)$, and we can also provide the converged $D = 6$ result $-0.62774(1)$.
All our energies as well as our quantum Monte Carlo reference values corresponding to Fig.~\ref{fig:HeisEnergies} are collected in Appendix \ref{app:pepsE}.
Our $D = 4$ energy per site is already lower than the best values reported for the wave function ansatzes (block) sequentially generated states ($-0.61713$) \cite{BanulsSGS}, entangled-plaquette states ($-0.6258(1)$) \cite{MezzacapoEPS}, and string bond states ($-0.6225$) \cite{SfondriniSBS}, to which we can directly compare because they also considered finite systems with open boundary conditions.
For infinite systems, the iPEPS ansatz attains slightly better energy precisions between $10^{-3}$ and $10^{-4}$ for $D = 4$ to $6$, as reported in Ref.~\cite{Stoudenmire2DDMRG}.
And for large finite cylinders, the best DMRG results are also more accurate:
Reference \cite{White2DHeis} analyzes the Heisenberg model on a cylinder with a constant staggered magnetic field on the boundaries and, by making use of $S^{Z}$ symmetry in the algorithm, reaches an energy accuracy of $10^{-4}$ on a $20 \times 10$ lattice.

In order to check the accuracy of the ground state approximation, we evaluated also non-local observables.
In particular, we computed the correlator $\langle \vec{S}_{l} \cdot \vec{S}_{l+x} \rangle$, in the center of the lattice for two sites separated by a distance $x$, either along the diagonal or along the same column.
We checked explicitly that the correlators of the converged PEPS along the diagonal and vertical direction are quantitatively very similar.
This feature is obviously due to the PEPS ansatz and would be harder to reproduce e.g.\ with MPS in two dimensions.
The precision of our considered spin-spin correlator $\langle \vec{S}_{l} \cdot \vec{S}_{l+x} \rangle$ also indicates the precision that can be expected for the order parameter $M_{\mathrm{stag}}^{2} := \frac{1}{N^{2}} \sum_{l, m = 1}^{N} (-1)^{l+m} \langle \vec{S}_{l} \cdot \vec{S}_{m} \rangle$.
Since the former quantity, being dependent on the distance $x$, provides more information than the latter quantity, being just a single number, we focus here on the spin-spin correlator.

\begin{figure}
\centering
\includegraphics[width=0.443\textwidth]{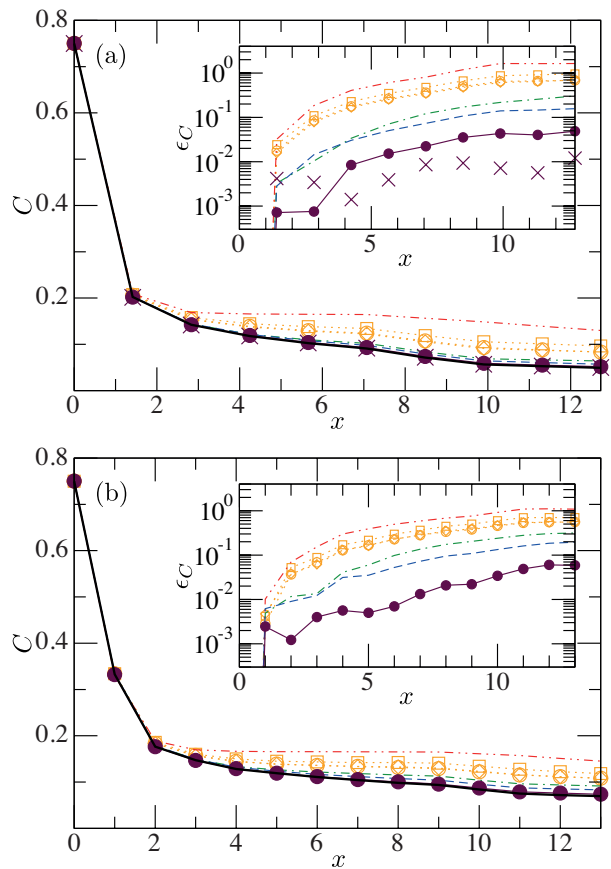}
\caption{\label{fig:HeisSpinCorr}
                Spin correlations $C(x) := | \langle \vec{S}_{l} \cdot \vec{S}_{l+x} \rangle |$ (main) and relative error $\epsilon_{C}(x) := |C(x)-C_{0}(x)|/|C_{0}(x)|$ (inset), with the exact values $C_{0}(x)$ (thick line) from the ALPS library \cite{ALPS1, ALPS2, ALPS3},
                for two sites separated by distance $x$ along the diagonal in the center of $10 \times 10$ PEPS (a) and along the vertical in the center of $14 \times 14$ PEPS (b).
                We consider PEPS Heisenberg ground state approximations from the FU with $D=2$ (dash-double-dotted), $4$ (dash-dotted), $5$ (dashed), $6$ (filled circles), and $7$ (crosses),
                and from the SU with $D=4$ (squares), $6$ (diamonds), and $8$ (open circles).
               }
\end{figure}

The results for the diagonal correlators in $10 \times 10$ PEPS are shown in Fig.~\ref{fig:HeisSpinCorr} (a), and Fig.~\ref{fig:HeisSpinCorr} (b) displays the vertical correlators for $14 \times 14$ PEPS.
We observe that the FU converges quickly to the true correlator with increasing bond dimension.
Although for fixed $D$ the error grows with the distance $x$, for fixed $x$ it decreases fast with $D$.
In particular, if we consider the correlator at distance $x=L/2$, as commonly done for the construction of the thermodynamic value via finite size scaling, we read off $\epsilon_{C}^{D=6} \approx 0.01$ and $\epsilon_{C}^{D=7} \approx 0.003$ on the $10 \times 10$ lattice, and we find $\epsilon_{C}^{D=5} \approx 0.07$ and $\epsilon_{C}^{D=6} \approx 0.01$ on the $14 \times 14$ lattice.
As for the energy, the SU results saturate, and they get better when the system size is larger.

We want to compare our results for the spin-spin correlator to previous works.
The widely used iPEPS algorithms achieve a remarkably low relative energy error in the thermodynamic limit \cite{Stoudenmire2DDMRG} while their relative correlator error $\approx 0.1$ reported in Ref.~\cite{BaueriPEPS} for $D = 5$ is still rather high (although larger values of $D$ are accessible within iPEPS algorithms nowadays \cite{CorboztJ, CorbozSSM} by making use of symmetries \cite{VidalSymmTNS, BauerSymmPEPS}).
In Ref.~\cite{WangMCTNS} the SU was used together with Monte Carlo sampling to reach much larger bond dimensions, and their best accuracies obtained with $D = 16$ were $0.003(2)$ on a $8 \times 8$ lattice and $0.013(2)$ on a $16 \times 16$ lattice, assuming periodic boundary conditions.
We now attain the same precisions here on $10 \times 10$ and $14 \times 14$ lattices already with much smaller bond dimensions $D = 6$ and $7$.
Again, the best DMRG results are still more accurate:
Reference \cite{White2DHeis} reports an uncertainty of $0.0007$ for the observable $|\langle S^{Z} \rangle|$ in the center of a $20 \times 10$ cylinder with constant staggered magnetic fields on the boundaries.

\begin{figure}
\centering
\includegraphics[width=0.443\textwidth]{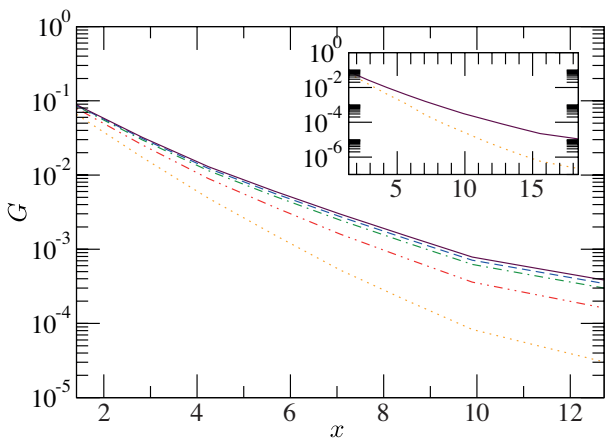}
\caption{\label{fig:HeisCorrFunc}
                Correlation function $G(x) := \langle \vec{S}_{l} \cdot \vec{S}_{l+x} \rangle - \langle \vec{S}_{l} \rangle \cdot \langle \vec{S}_{l+x} \rangle$ for two sites separated by distance $x$ along the diagonal in the center of
                $10 \times 10$ (main) and $14 \times 14$ (inset) PEPS Heisenberg ground state approximations of bond dimension $D=4$.
                We compare SU (dotted), CU$_{1}$ (dash-double-dotted), CU$_{2}$ (dash-dotted), CU$_{3}$ (dashed), and FU (solid).
               }
\end{figure}

We can try to understand the characteristics of the SU and the FU results with the help of the environment approximation used in their tensor updates.
As we have argued in Ref.\ \cite{Lubasch2}, SU and FU represent special cases of a unifying CU$_{\delta}$: the SU is equivalent to clusters of size $\delta=0$ in the tensor update, while the FU corresponds to the largest possible cluster size $\delta=L-1$.
Here we showed in Sec.\ \ref{sec:impr}, that the cluster contraction error as a function of the cluster size behaves like the correlation function of the considered PEPS, such that states with short correlation lengths can be accurately contracted by means of small clusters.
It is then reasonable to expect that the cluster size $\delta$ used in CU$_{\delta}$ limits the finally achievable correlation length.
We address this question on a $10 \times 10$ lattice with $D=4$ in the main part of Fig.~\ref{fig:HeisCorrFunc}.
Indeed, the correlation function decays slower when larger clusters are used in the CU.

Moreover, we can gather from Fig.~\ref{fig:HeisCorrFunc} that the correlation functions for system size $14 \times 14$ from SU as well as FU decay faster than the corresponding ones for system size $10 \times 10$, while Fig.~\ref{fig:HeisEnergies} shows that a higher energy accuracy is attained on the larger lattice.
This indicates that, for the finite systems with open boundary conditions considered here, the true correlation length of the Heisenberg model slightly decreases with growing lattice size.
In the context of the SU, this would explain why the SU results of Figs.~\ref{fig:HeisEnergies} and \ref{fig:HeisSpinCorr} are better on larger lattices.
And in the context of the FU, this would explain our numerical observation that the convergence of energies and spin-spin correlators required smaller values of $D'$ for larger systems
\footnote{
To be precise, our FU energies and spin-spin correlators for the Heisenberg model were converged with $D' = 2D^{2}$ for all $D = 2$ to $6$ on the $14 \times 14$ lattice, i.e.\ our results did not change anymore when we further ran the FU with larger $D'$.
However, this convergence occurred only with $D' = 75$ for $D = 5$ and $D' = 126$ for $D = 6$ on the $10 \times 10$ lattice.
}:
A smaller correlation length can be captured with a smaller cluster size $\delta$ in the CU$_{\delta}$ and the contraction precision achieved with such $\delta$ can equally be obtained by the full contraction, used in the FU, with correspondingly smaller value of $D'$ (see Figs.\ 12 and 13 in Ref.~\cite{Lubasch2}).

\subsection{Quantum Ising model}

We have also applied our finite PEPS algorithms to the quantum Ising model with transverse field, \mbox{$\hat{H} = -\sum_{\langle l, m \rangle} \sigma_{l}^{Z} \sigma_{m}^{Z} -B \sum_{l} \sigma_{l}^{X}$}.
This Hamiltonian features a quantum phase transition in the thermodynamic limit, and its critical point $B_{c} \approx 3.044$ and exponent $\beta \approx 0.327$ are known very accurately thanks to finite size scaling with quantum Monte Carlo \cite{Bloete}.
Since iPEPS have already very successfully demonstrated the adequacy of the PEPS ansatz for the quantum Ising model even at criticality \cite{CiraciPEPS, OrusCTM} (Ref.~\cite{OrusCTM} reports $B_{c} \approx 3.04$ and $\beta \approx 0.328$), we present our results here and in Appendix \ref{app:pepsE} just for benchmark purposes, e.g.\ to enable a comparison with another PEPS implementation or with another wave function ansatz.
We thus consider here only few different values of the magnetic field around $B = 3$ and run our computations only for the two system sizes $11 \times 11$ and $21 \times 21$.
For each value of $B$, we converge the imaginary time evolution independently, using time steps $\tau=10^{-2}$, $10^{-3}$, and $10^{-4}$.

\begin{figure}
\centering
\includegraphics[width=0.455\textwidth]{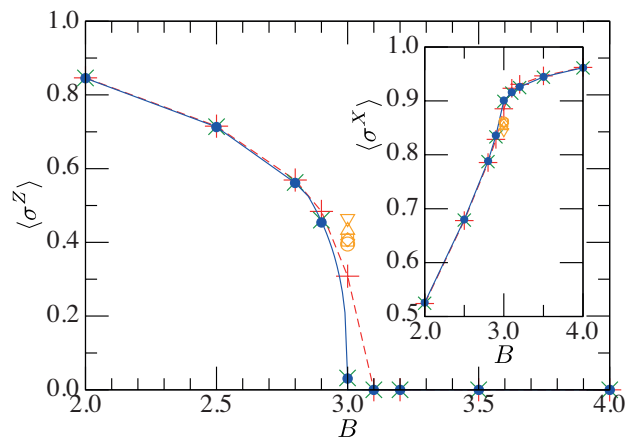}
\caption{\label{fig:IsingOp}
                Observables $\langle \sigma^{Z} \rangle$ (main) and $\langle \sigma^{X} \rangle$ (inset) evaluated in the center of $21 \times 21$ PEPS Ising ground state approximations from the FU with $D=2$ (plusses), $3$ (crosses), and $4$ (filled circles),
                where the $D=4$ are basically on top of the $D=3$ results.
                Open symbols show the SU at $B=3.0$, for $D=2$ (down-triangles), $4$ (up-triangles), $5$ (squares), $6$ (diamonds), and $7$ (open circles).
                We interpolate the FU $D=4$ $\langle \sigma^{Z} \rangle$ results between $B=2.85$ and $B_{0}:=3.000035$ with $|B-B_{0}|^{0.34}$.
               }
\end{figure}

Figure \ref{fig:IsingOp} shows the order parameter evaluated in the center of $21 \times 21$ PEPS from the FU for several points in the phase diagram.
Without performing a finite size scaling, we can already extract estimates of the critical point $B_{c} \approx 3.0$ and exponent $\beta \approx 0.34$ from this finite system, which are close to the iPEPS results \cite{CiraciPEPS, OrusCTM}.
We conclude that this lattice is already large enough to display features similar to iPEPS.

\begin{figure}
\centering
\includegraphics[width=0.449\textwidth]{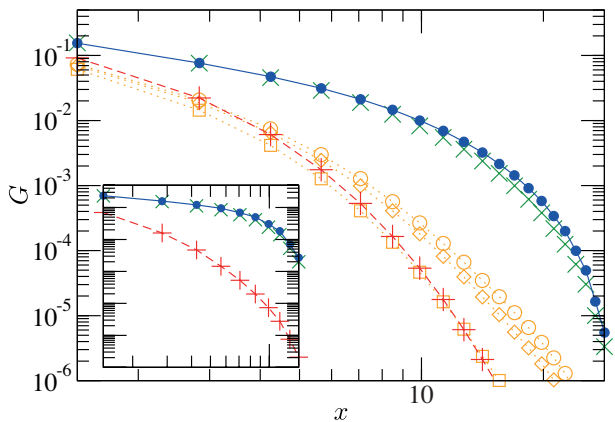}
\caption{\label{fig:IsingCorrFunc}
                Correlation function $G(x) := \langle \sigma_{l}^{Z} \sigma_{l+x}^{Z} \rangle - \langle \sigma_{l}^{Z} \rangle \langle \sigma_{l+x}^{Z} \rangle$ for two sites separated by distance $x$ along the diagonal in the center of
                PEPS ground state approximations of the Ising model on a $21 \times 21$ lattice with transverse field $B=3.0$ (main) and on a $11 \times 11$ lattice with $B=2.8$ (inset).
                The results were obtained with the FU using bond dimension $D=2$ (plusses), $3$ (crosses), and $4$ (filled circles), and with the SU using $D=4$ (squares), $6$ (diamonds), and $7$ (open circles).
               }
\end{figure}

For comparison, we also present SU results at $B=3$.
As expected from our previous analysis, the SU does not work well there, where the correlation length should be large.
Figure \ref{fig:IsingCorrFunc} shows that the FU can indeed generate PEPS with larger correlation lengths.
While a least squares fit gives a correlation length for the FU $D=4$ PEPS $\zeta_{FU}^{D=4} \approx 2.6$, it reveals for the SU $D=7$ PEPS only $\zeta_{SU}^{D=7} \approx 1.2$.
The inset of Fig.~\ref{fig:IsingCorrFunc} demonstrates the largest correlation length $\zeta_{FU}^{D=4} \approx 4.3$ for the $11 \times 11$ lattice.
Notice that, here, we have not performed such an extensive convergence analysis with $D'$ as we have done before for the Heisenberg Hamiltonian
\footnote{For the quantum Ising model and all considered transverse fields $B$, on the $11 \times 11$ lattice, we ran the FU with $D' = 8$ for $D = 2$, $D' = 54$ for $D = 3$, and $D' = 128$ for $D = 4$, while on the $21 \times 21$ lattice, we ran the FU with $D' = 8$ for $D = 2$, $D' = 36$ for $D = 3$, and $D' = 128$ for $D = 4$.}.
Nevertheless, we want to emphasize that our correlation functions for the $21 \times 21$ lattice are in perfect agreement with the best iPEPS results \cite{OrusCTM}.

Remarkably, long correlation lengths can be analyzed, i.e.\ large clusters can be contracted, with very high accuracy in the framework of PEPS.
This constitutes clear evidence for the power of general boundary MPO.
They can capture the correlations of a large cluster size $\delta$ with a boundary bond dimension much smaller than the one needed for the exact contraction, $D'=D^{2\delta}$.

\section{Conclusions}
\label{sec:conclusions}

In this paper we have reviewed various aspects that need to be taken into account in the implementation of efficient state-of-the-art finite PEPS algorithms.
Within the two main parts of PEPS algorithms, namely the environment approximation and the tensor update, we have analyzed algorithmic strategies that improve the efficiency and stability of the procedures, and the physical properties of the solution.

The environment approximation has decisive influence on the precision of the final PEPS of an imaginary time evolution, and is equally crucial for the computation of expectation values.
We have shown how the accuracy of the cluster strategy \cite{Lubasch2}, which allows for a natural trade-off between precision and computational cost of the environment, is fundamentally connected to the correlation length of the state.
Additionally, we have demonstrated that it is possible to make use of purification MPO in order to ensure a positive environment approximation, and that this overcomes the limitations of the Single-Layer algorithm \cite{PizornSL}.
The numerical techniques analyzed in this problem can straightforwardly be applied to the Cluster Update \cite{Lubasch2} and to the Full Update, but also to other scenarios where a positive MPO is required, e.g.\ to describe the mixed state of a one-dimensional system.

Not only the environment approximation, but also the method chosen for the tensor update affects the cost and stability of the routines.
We have proposed an update scheme that is more efficient and better conditioned than the one from the original algorithm \cite{CiracOriginalPEPS, CiracReview1}.
By restricting the variational parameters to the reduced tensor, the update is drastically accelerated.
For both the reduced and the full tensor, we have formulated gauge fixings that significantly improve the conditioning.
These gauge fixings, additionally, when combined with a cheap SVD, constitute a promising simplified but fast tensor update procedure.

Finally, we have combined the ingredients discussed above in an efficient implementation of finite PEPS imaginary time evolution, capable of dealing with large systems and bond dimensions.
In particular we have opted for the sequential application of Trotter gates, using general boundary MPO in the contraction of the full environment
\footnote{Notice that clusters could be beneficial in a parallel implementation as discussed in Ref.\ \cite{Lubasch2}.},
and restricting the update to the reduced tensors.
To benchmark the performance of finite PEPS and to quantitatively assess the algorithmic properties, we have applied the code to the ground state search for the Heisenberg and the quantum Ising model.

We have presented ground state calculations for system sizes up to $21 \times 21$ and bond dimensions up to $D \approx 7$, $8$.
Our results demonstrate the adequacy of the PEPS ansatz for the description of strongly correlated quantum many-body systems, with energy and order parameter converging fast with increasing bond dimension, when they were obtained with the Full Update.
In that case, thanks to the algorithmic improvements developed in this article, we have been able to achieve precisions of the spin-spin correlator in the Heisenberg model using bond dimensions $D = 6$ and $7$ that previously had only been attained using a much larger $D = 16$ in Ref.~\cite{WangMCTNS}.
Our analysis of a $21 \times 21$ quantum Ising model gave, already without finite size scaling, critical point, critical exponent and correlation functions in good agreement with the iPEPS results \cite{CiraciPEPS, OrusCTM}.

The Simple Update \cite{XiangSU} and the Cluster Update \cite{Lubasch2} using small cluster sizes, while ensuring a less costly environment and thus being able to deal with larger bond dimensions, do not produce the best ground state approximation for a certain value of $D$, and, in particular, give rise to PEPS with limited correlation length, which is especially relevant for strongly correlated systems as e.g.\ the Ising model close to criticality.
This makes clear that the largest bond dimension attained is not the significant measure of the power of a PEPS algorithm.

By reaching system sizes typically considered for finite size scaling, we have given evidence that finite PEPS, when all algorithmic details are taken into account, offer a feasible unbiased alternative to their infinite counterpart iPEPS \cite{CiraciPEPS}.
On the other hand, the algorithmic methods proposed here can also be applied to iPEPS, and the feasibility of large finite PEPS demonstrated here suggests that large unit cells are possible in iPEPS, such that their potential bias due to a finite unit cell can be well analyzed by systematically increasing the unit cell from small to very large size.

Our analysis has been carried out with a generic implementation of PEPS algorithms, so that one can expect that adapting the methods to the specific properties of a certain problem will further enhance the performance.
A particularly promising next step is to incorporate the symmetries of the considered Hamiltonian in the tensors \cite{VidalSymmTNS, BauerSymmPEPS}, a key element of ground-breaking two-dimensional DMRG studies, such as Refs.~\cite{WhiteKagome, SchollwoeckKagome}, and of seminal iPEPS calculations, such as Refs.~\cite{CorboztJ, CorbozSSM}.

\acknowledgements
M. L. thanks B. Bauer, P. Corboz, G. De las Cuevas, I. Pi\v{z}orn, and L. Wang for discussions.
He wants to particularly thank S. Iblisdir, V. Murg, R. Or\'{u}s, and M. Rizzi for providing key insights into algorithmic aspects.
This project was partly funded by the EU through SIQS grant (FP7 600645) and the DFG (NIM cluster of excellence).
All authors thank the Pedro Pascual Benasque Center for Science (CCBPP), where part of this work was carried out.

\appendix

\section{Purification approximations}
\label{app:puri}

Approximating the boundary by a purification MPO, as described in Sec.\ \ref{subsec:posenv}, requires the solution of nonlinear equations for each tensor $A_{l}$.
Different algorithms can be used for this purpose, and we have tried and compared three methods.

\begin{itemize}

\item Linearization:
Instead of solving the equations for the product $A_{l} A_{l}^{*}$, we solve them for $A_{l} B_{l}$, treating $A_{l}$ and $A_{l}^{*}$ as independent tensors.
In order to achieve convergence, the change of the tensor in each iteration needs to be small, and hence we construct the solution of the $i$th iteration according to $A_{l}^{(i)} = (1-\alpha) A_{l}^{(i-1)} + \alpha A_{l}$, where $A_{l}$ solves the linearized equations of the previous iteration and is added to the previous solution $A_{l}^{(i-1)}$ with a weight $\alpha$.
The latter parameter must be chosen small enough to guarantee a decreasing cost function, and large enough to avoid unnecessarily long convergence times
\footnote{We found that the optimization worked well if the initial value was $\alpha = 0.01$ and was multiplied by $0.8$ whenever the cost function increased.}.
The construction of the individual parts of the linear equations has the leading cost $\mathcal{O}(dD^{6}D''^{4})+\mathcal{O}(D^{4}D''^{6})$.
Because we cannot impose a kind of canonical form that gives a trivial norm matrix $N_{l} = 1$, we have to explicitly contract its tensor network, which contributes a cost $\mathcal{O}(d'^{2}D^{2}D''^{6})$.
Finally, computing the pseudoinverse to solve the linear equations requires $\mathcal{O}(d'^{3}D^{3}D''^{6})$ operations, which typically represent the dominant cost when $d' \geq D$.

\item Conjugate gradient:
We employ a canned routine
\footnote{For conjugate gradient minimization we use nag\_opt\_conj\_grad from the NAG library \cite{NAG}.}
that comprises a conjugate gradient method with line minimization.
It has the lowest computational cost, as it only requires the computation of the cost function and its gradient with respect to a single tensor, which can be obtained with $\mathcal{O}(dD^{6}D''^{4})+\mathcal{O}(D^{4}D''^{6})$ operations.

\item Newton method:
It approaches a root of the gradient by iterating $\mathcal{H}_{l}^{(i-1)}(\vec{A}_{l}^{(i)}-\vec{A}_{l}^{(i-1)}) = -\vec{\mathcal{G}}_{l}^{(i-1)}$, where $\mathcal{H}_{l}^{(i-1)}$ denotes the Hessian matrix and $\vec{\mathcal{G}}_{l}^{(i-1)}$ the gradient of the cost function with respect to the tensor components at site $l$, evaluated with the solution $A_{l}^{(i-1)}$ of the previous iteration
\footnote{
We observed that, occasionally, the Hessian matrix had many negative eigenvalues and then it was crucial to include only the positive ones (above a certain cutoff) in the construction of its pseudoinverse.
}.
The Newton method has the advantage that the step width is naturally given, in contrast to the linearized equations where $\alpha$ needs to be chosen heuristically, and in contrast to the conjugate gradient routine where it is determined via line search.
In addition to the parts of the conjugate gradient algorithm, the Newton method needs the Hessian matrix of the cost function, which contributes $\mathcal{O}(d'^{2}D^{2}D''^{6})$ to the cost, and its pseudoinverse, determined by $\mathcal{O}(d'^{3}D^{3}D''^{6})$ operations, such that the leading cost $\mathcal{O}(dD^{6}D''^{4})+\mathcal{O}(D^{4}D''^{6})+\mathcal{O}(d'^{3}D^{3}D''^{6})$ is the same as for the linearized problem.

\end{itemize}

To compare the different alternatives, we benchmarked their performance in the search for an optimal purification with fixed $D''=2$ and varying $d'$, given a reference purification with $D''=4$ and $d'=4$.
The latter was constructed by taking two rows from one edge of a PEPS norm TN, for several $11 \times 11$ $D=2$ SU ground state approximations of the Ising model at various magnetic fields.
As a general rule, the initial tensors for the search with incremented purification bond $d'+1$ were chosen as the previous solution for $d'$ where the extra elements were filled with uniformly distributed random numbers.
From the three considered algorithms, the Newton method performed best.
It converged reliably for all $d'$ within few local updates per tensor.

All the methods benefit from initial variables that are already close to the final solution.
A sensible numerical approach to purification approximations can then be implemented in two steps: firstly, the computation of the optimal purification via the SL algorithm \cite{PizornSL}, and secondly, the further optimization of that purification via the Newton method.

\section{Gauge fixing for the full tensors}
\label{app:gaug}

\begin{figure}
\centering
\includegraphics[width=0.414\textwidth]{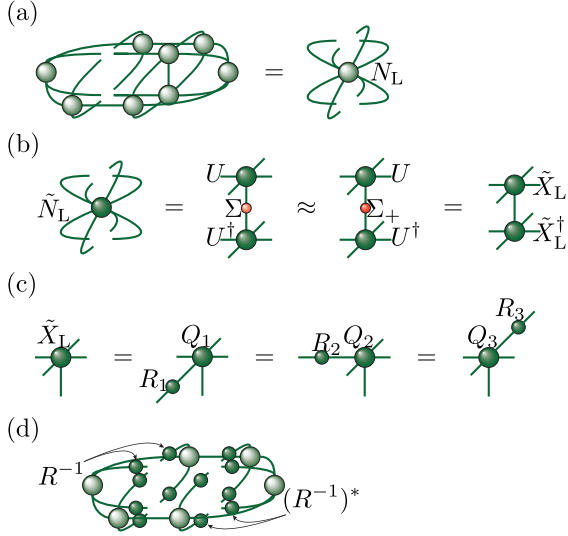}
\caption{\label{fig:GaugeEnvFullTensorPair}
                Gauge fixing on the environment tensors of the full tensor pair, when the environment is non-separable.
                (a) The environment tensor $N_{\mathrm{L}}$ of the left full tensor.
                (b) We determine the positive approximant for the environment of the left full tensor via a diagonalization of the hermitian approximant
                      $\tilde{N}_{\mathrm{L}} := (N_{\mathrm{L}}+N_{\mathrm{L}}^{\dag})/2 = U \Sigma U^{\dag}$, in which, then, the negative eigenvalues are discarded in $\Sigma_{+}$, and, finally, the environment is written
                      as $\tilde{X}_{\mathrm{L}} \tilde{X}_{\mathrm{L}}^{\dag}$ in terms of its square root $\tilde{X}_{\mathrm{L}} := U\sqrt{\Sigma_{+}}$.
                (c) We perform three independent QR decompositions on $\tilde{X}_{\mathrm{L}}$.
                (d) After equally having carried out the previous steps (a) to (c) with the right full tensor, we have six different matrices $R$.
                      Their inverses are contracted with the corresponding tensors of the boundary MPO.
               }
\end{figure}

When the environment is non-separable, and the update of the full tensors is considered, gauge transformations can be efficiently computed in such a way that an eigendecomposition of the $D^{6} \times D^{6}$ dimensional norm environment of the pair is not necessary.
Instead, the $D^{4} \times D^{4}$ dimensional environments of the left ($N_{\mathrm{L}}$) and right ($N_{\mathrm{R}}$) tensor are independently computed (Fig.~\ref{fig:GaugeEnvFullTensorPair} (a)) and replaced by their positive approximants (Fig.~\ref{fig:GaugeEnvFullTensorPair} (b)) like in previous cases.
Their square roots are used to obtain the desired gauge transformations for each of the virtual bonds of the pair (Fig.~\ref{fig:GaugeEnvFullTensorPair} (c)).
On each virtual bond we then insert the corresponding product $R^{-1} R$ and absorb the $R$ matrices in the full tensors and their inverses in the environment (Fig.~\ref{fig:GaugeEnvFullTensorPair} (d)).

\begin{figure}
\centering
\includegraphics[width=0.471\textwidth]{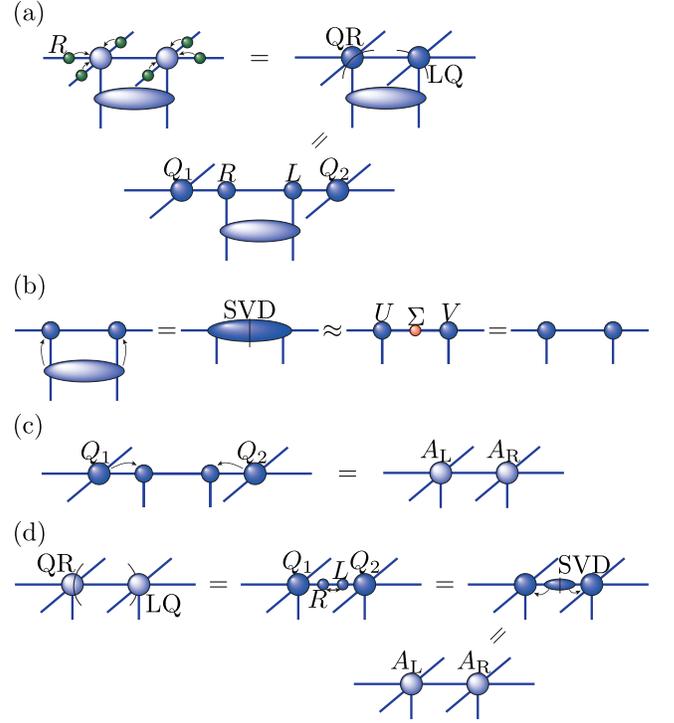}
\caption{\label{fig:UpdateFullTensorPair}
                Like in the reduced case of Fig.~\ref{fig:UpdateRedTensorPair}, the update of the full tensor pair also consists of the three stages initialization, optimization, and final form.
                The optimization is the standard ALS sweeping, in which each full tensor is gauged after its update in the standard way, i.e.\ the left tensor is QR decomposed along its right virtual bond and the right
                tensor is LQ decomposed along its left virtual bond.
                (a) Initialization I: Firstly, we contract the gauge transformations from Fig.~\ref{fig:GaugeEnvFullTensorPair} with the full tensors, and split off their reduced parts.
                (b) Initialization II: Secondly, we construct new reduced tensors from a SVD on the tensor pair and the Trotter gate, equally sharing the $D$ largest singular values between the left and right tensor.
                (c) Initialization III: We recover the full tensors $A_{\mathrm{L}}$ and $A_{\mathrm{R}}$, which are now the initial tensors for the optimization via ALS sweeping.
                (d) Final form: After convergence of the sweeping, we put the two tensors on the same footing.
               }
\end{figure}

The update of the full tensor pair proceeds in the way explained in Fig.~\ref{fig:UpdateFullTensorPair}, analogously to the reduced tensor update.
As in the latter context, if the environment is separable, the tensor initialization Fig.~\ref{fig:UpdateFullTensorPair} (a) to (c) already minimizes the cost function, while, if the environment is close to separable, we can expect a significant decrease of the cost function.
In general, we can anticipate good starting tensors for the following ALS sweeping.

Table \ref{tab:CondnumsFullTensor} contains typical condition numbers of the norm matrix in the full tensor update without our gauge fixing, using only the positive approximant, and with our gauge fixing.
Our gauge fixing improves the condition number drastically.

\begin{table}[!t]
\centering
 \begin{tabular}{ c c c }
  \hline \hline
  Model               & Positive approximant            & Gauge fixing \\ \hline
  $B=1.0$ Ising & $(2 \pm 3) \cdot 10^{9}$       & $1.3 \pm 0.2$ \\
  $B=3.0$ Ising & $(2 \pm 2) \cdot 10^{4}$       & $2.9 \pm 0.7$ \\
  Heisenberg    & $(1.3 \pm 0.8) \cdot 10^{3}$ & $1.15 \pm 0.05$ \\
  \hline \hline
 \end{tabular}
\caption{\label{tab:CondnumsFullTensor}
               We show the mean condition number of the norm matrix with its standard deviation in the full tensor update, for $D=2$ and the setting of Tab.~\ref{tab:CondnumsRedTensor}.
              }
\end{table}

\newpage

\section{Finite PEPS energies}
\label{app:pepsE}

Here we collect some precise energy values obtained with the PEPS ground state approximations considered in this paper.

In the case of the Heisenberg model, we compare our results to energies from the quantum Monte Carlo loop algorithm of the ALPS library \cite{ALPS1, ALPS2, ALPS3}, summarized in Tab.~\ref{tab:HeisQMC}.
The presented values and errors correspond to temperature $T=10^{-4}$, and they agree with the ones corresponding to $T=10^{-3}$ within the error bars.

\begin{table}[!h]
\centering
 \begin{tabular}{ c c c c }
  \hline \hline
  $10 \times 10$ & $14 \times 14$ & $16 \times 16$ & $20 \times 20$ \\ \hline
   -0.628656(2) & -0.639939(2) & -0.643531(2) & -0.648607(1) \\
  \hline \hline
 \end{tabular}
\caption{\label{tab:HeisQMC}
               Energy per site of the Heisenberg model on square lattices of various sizes from quantum Monte Carlo, computed with the ALPS library \cite{ALPS1, ALPS2, ALPS3}.
              }
\end{table}

\begin{table}[!h]
\centering
 \begin{tabular}{ c c c }
  \hline \hline
  $D$ & $10 \times 10$ & $14 \times 14$ \\ \hline
    2    &   -0.61310(2)     &  -0.62631(1)     \\
    3    &   -0.61999(1)     &  -0.63246(1)     \\
    4    &   -0.62637(2)     &  -0.63832(3)     \\
    5    &   -0.62739(1)     &  -0.63901(1)     \\
    6    &   -0.62774(1)     &  -0.63930(1)     \\
  \hline \hline
 \end{tabular}
\caption{\label{tab:HeisFU}
               Energy per site of PEPS Heisenberg ground state approximations from the FU.
              }
\end{table}

\begin{table}[!h]
\centering
 \begin{tabular}{ c c c c c }
  \hline \hline
  $D$ & $10 \times 10$ & $14 \times 14$ & $16 \times 16$ & $20 \times 20$ \\ \hline
    2    &   -0.61281(1)    &   -0.62115(1)     &   -0.62719(2)    &   -0.63519(2)     \\
    3    &   -0.61846(2)    &   -0.62977(1)     &   -0.63433(1)    &   -0.64056(2)     \\
    4    &   -0.62382(1)    &   -0.63587(1)     &   -0.63985(1)    &   -0.64549(2)     \\
    5    &   -0.62520(2)    &   -0.63713(2)     &   -0.64106(1)    &   -0.64659(2)     \\
    6    &   -0.62541(2)    &   -0.63738(2)     &   -0.64129(2)    &   -0.64676(2)     \\
  \hline \hline
 \end{tabular}
\caption{\label{tab:HeisSU}
               Energy per site of PEPS Heisenberg ground state approximations from the SU.
              }
\end{table}

\begin{table}[!h]
\centering
 \begin{tabular}{ c c c c }
  \hline \hline
  $D$ & $2.0$            & $2.5$            & $2.8$            \\ \hline
    2    & -2.40075(1) & -2.74230(2) & -2.98947(5) \\
    3    & -2.40076(1) & -2.74243(1) & -2.99094(2) \\
    4    & -2.40076(1) & -2.74243(1) & -2.99099(1) \\
  \hline \hline
 \end{tabular}
 \begin{tabular}{ c c c c }
  \hline \hline
  $D$ & $2.9$            & $3.0$            & $3.1$           \\ \hline
    2    & -3.07945(5) & -3.17128(4) & -3.26400(4) \\
    3    & -3.08071(1) & -3.17210(1) & -3.26457(1) \\
    4    & -3.08073(1) & -3.17210(1) & -3.26457(1) \\
  \hline \hline
 \end{tabular}
 \begin{tabular}{ c c c c }
  \hline \hline
  $D$ & $3.2$            & $3.5$            & $4.0$           \\ \hline
    2    & -3.35744(4) & -3.64097(3) & -4.12064(2) \\
    3    & -3.35785(1) & -3.64116(1) & -4.12071(1) \\
    4    & -3.35785(1) & -3.64116(1) & -4.12071(1) \\
  \hline \hline
 \end{tabular}
\caption{\label{tab:IsingFUL11}
               Energy per site of $11 \times 11$ PEPS Ising ground state approximations from the FU for different transverse fields $B$.
              }
\end{table}

\begin{table}[!h]
\centering
 \begin{tabular}{ c c c c }
  \hline \hline
  $D$ & $2.0$            & $2.5$            & $2.8$            \\ \hline
    2    & -2.45219(1) & -2.77340(2) & -3.00705(4) \\
    3    & -2.45219(1) & -2.77346(1) & -3.00737(1) \\
    4    & -2.45219(1) & -2.77346(1) & -3.00737(1) \\
  \hline \hline
 \end{tabular}
 \begin{tabular}{ c c c c }
  \hline \hline
  $D$ & $2.9$            & $3.0$            & $3.1$            \\ \hline
    2    & -3.09228(5) & -3.18128(6) & -3.27326(4) \\
    3    & -3.09287(1) & -3.18242(1) & -3.27406(1) \\
    4    & -3.09287(1) & -3.18243(1) & -3.27406(1) \\
  \hline \hline
 \end{tabular}
 \begin{tabular}{ c c c c }
  \hline \hline
  $D$ & $3.2$            & $3.5$            & $4.0$            \\ \hline
    2    & -3.36617(4) & -3.64849(3) & -4.12685(2) \\
    3    & -3.36672(1) & -3.64873(1) & -4.12694(1) \\
    4    & -3.36672(1) & -3.64873(1) & -4.12694(1) \\
  \hline \hline
 \end{tabular}
\caption{\label{tab:IsingFUL21}
               Energy per site of $21 \times 21$ PEPS Ising ground state approximations from the FU for different transverse fields $B$.
              }
\end{table}

\begin{table}[!h]
\centering
 \begin{tabular}{ c c }
  \hline \hline
  $D$ & 3.0              \\ \hline
    2    & -3.1792(4) \\
    3    & -3.1806(4) \\
    4    & -3.1807(4) \\
    5    & -3.1812(5) \\
    6    & -3.1812(4) \\
    7    & -3.1814(5) \\
  \hline \hline
 \end{tabular}
\caption{\label{tab:IsingSU}
               Energy per site of $21 \times 21$ PEPS Ising ground state approximations from the SU for transverse field $B=3.0$.
              }
\end{table}

\end{document}